\documentclass[doublespacing]{bmcart}

\usepackage{amsthm,amsmath}
\usepackage[mathletters]{ucs}
\usepackage[utf8]{inputenc} 
\usepackage{booktabs}
\usepackage{tabularx}
\usepackage{graphicx}
\usepackage{adjustbox}
\usepackage{textcomp}
\usepackage{mathtools}
\usepackage[flushleft]{threeparttable}
\usepackage{setspace}
\usepackage{hyperref}
\usepackage{color,soul}
\usepackage{float}
\usepackage[T1]{fontenc}

\startlocaldefs
\endlocaldefs

\begin{document}

\begin{frontmatter}

\begin{fmbox}
\dochead{Research}

\title{Are Travel Bans the Answer to Stopping the Spread of COVID-19 Variants? Lessons from a Multi-Country SIR Model}

\author[
  addressref={aff1},                   
  email={frederic.docquier@liser.lu}   
]{\inits{}\fnm{Frédéric} \snm{Docquier}}
\author[
  addressref={aff2},
  email={nicolas.golenvaux@uclouvain.be}
]{\inits{}\fnm{Nicolas} \snm{Golenvaux}}
\author[
  addressref={aff2},
  corref={aff2},                       
  email={pierre.schaus@uclouvain.be}
]{\inits{}\fnm{Pierre} \snm{Schaus}}

\address[id=aff1]{
  \orgdiv{Crossing Borders},             
  \orgname{Luxembourg Institute for Socio-Economic Research (LISER)},          
  \city{Esch-sur-Alzette},                    
  \cny{Luxembourg}                            
}
\address[id=aff2]{%
  \orgdiv{Institute for Information and Communication Technologies, Electronics and Applied Mathematics (ICTEAM)},
  \orgname{Universit\'{e} catholique de Louvain},
  \city{Louvain-la-Neuve},
  \cny{Belgium}
}

\end{fmbox}

\begin{abstractbox}

\begin{abstract} 
\parttitle{Background} 
Detections of mutations of the SARS-CoV-2 virus gave rise to new packages of interventions. Among them, international travel restrictions have been one of the fastest and most visible responses to limit the spread of the variants. While inducing large economic losses, the epidemiological consequences of such travel restrictions are highly uncertain. They may be poorly effective when the new highly transmissible strain of the virus already circulates in many regions. Assessing the effectiveness of travel bans is difficult given the paucity of data on daily cross-border mobility and on existing variant circulation. The question is topical and timely as the new omicron variant -- classified as a variant of concern by WHO -- has been detected in Southern Africa, and perceived as (potentially) more contagious than previous strains. In this study, we develop a multi-country compartmental model of the SIR type. We use it to simulate the spread of a new variant across European countries, and to assess the effectiveness of unilateral and multilateral travel bans.

\parttitle{Results} 
Multilateral travel bans do not buy much time, by increasing the time until the infection curve peaks by a few weeks at best. This can be achieved with drastic travel bans only, and this fails to stop the propagation when the virus is already circulating in the country, or in regions not included in the travel bans. 

\parttitle{Conclusion} 
Given their huge economic and freedom-killing consequences, travel bans have negligible effects on the timing and severity of the infection response. Managing new waves of COVID-19 with local sanitary measures applicable to cross-border movers is the most effective option. It induces better epidemiological outcomes and smaller economic cost for all parties concerned. 

\end{abstract}

\begin{keyword}
\kwd{Cross-border mobility}
\kwd{Covid-19}
\kwd{SIR}
\end{keyword}


\end{abstractbox}

%

\end{frontmatter}




\section*{Background}

The overall pattern of the COVID-19 pandemic has been a series of ups and downs in infection rates. Virus mutations partly explain surges in the number of new cases, as they affect transmissibility, disease severity and the effectiveness of vaccines. In particular, the third wave that began in the summer 2021 has been closely associated with the spread of the contagious delta variant of SARS-CoV-2.\footnote{ During the 8th meeting of the IHR Emergency Committee on COVID-19 on July 14, Tedros Adhanom Ghebreyesus, the head of WHO, said: ``The Delta variant is one of the main drivers of the current increase in transmission, fueled by increased social mixing and mobility, and inconsistent use of proven public health and social measures.'' \cite{adhanom_ghebreyesus_who_nodate}.} Earlier, between December 2020 and January 2021, four other variants were classified as ``variants of concern'' (VOC) by the World Health Organization (WHO) -- the alpha (British), beta (South African), gamma (Brazilian) and delta (India) variants – and required new packages of sanitary measures \cite{noauthor_tracking_nodate}. Very recently, a new omicron variant was identified in South Africa (from a sample collected on November 9). Omicron has likely been circulating for a while and widely.  Although its transmissibility advantage is yet to be proven, it is perceived as re-infecting people more easily than other strains, and the WHO also classified it as a VOC.

Each variant and each wave called for a package of policy measures. Systematically, however, disrupting human mobility has been one of the fastest and most visible responses to limit the international spread of COVID-19 variants. At the early stages of the pandemic, WHO’s recommendation against ``any travel or trade restriction'' became a point of criticism. Most countries adopted some form of mobility restrictions -- e.g., travel bans, visa restrictions, border closures, among others -- with little reproach from the international community. Beginning of January 2021, many countries closed their borders with the UK after the alpha and beta variants were detected by British health experts. End of July 2021, the U.S. State Department issued a series of “do not travel” advisories for selected countries because of a rising infection rates in those countries, while European leaders enforced the EU Digital COVID-19 Certificate, or stricter testing and quarantining measures for non-vaccinated travelers. History is repeating itself as the omicron variant has quickly been detected in the UK, Germany, Czech Republic, Israel, Hong Kong and Belgium. Less than one week after its detection, the European Commission called on the EU Member States to introduce a general European flight ban to and from seven southern African countries. Immediately thereafter, Australia decided to close its borders for all foreigners, and many other countries (including the U.S., Canada, Japan and Israel) announced severe travel restrictions from Southern Africa.

It is quite probable that a wider range of cross-border restrictions have been adopted during the COVID-19 pandemic than in past disease outbreaks, which has induced large socioeconomic consequences. In the \textit{International Migration 2020 Highlights}, the United Nations wrote that ``the pandemic may have slowed the growth in the stock of international migrants by around two million by mid-2020, 27 per cent less than the growth expected since mid-2019,'' and emphasized the detrimental effect on existing vulnerabilities worldwide (e.g. trapped populations, reduced remittances, reduced return migration, longer visa processing time for refugees, etc.). The current bans on movements from Southern African will also have significant adverse effects on the economies, which traditionally welcome global tourists over the summer year-end period. Given these huge economic and freedom-killing consequences, travel bans should be based on evidence, not politics or fear. Existing literature \cite{priesemann2021, chen2020, lee2020, errett2020, chinazzi2020, asiktas2021, tian2020, kraemer2020} on these questions offers mixed results. On the one hand, by connecting any two points on the planet, there is legitimate reason to think that cross-border human mobility can contribute to a swifter and broader propagation of infectious diseases.\footnote{At the very early stage of the COVID-19 outbreak and in most countries, patient zero typically was a traveler who spent time in a high-prevalence area.} On the other hand, border controls and travel restrictions may be poorly effective when the new highly transmissible strains of the virus already circulate in virtually all regions. Whether the effect of travel bans on the spread of the virus is large or small is an empirical question, whose investigation requires data on infection/transmission rates and cross-border mobility. The latter are usually unavailable or imprecise.

In this paper, we develop and calibrate a multi-country compartmental model of the SIR type to shed light on the epidemiological implications of travel bans. We parameterize it to match the epidemiological context of the spread of a new and more contagious variant in Europe, accounting for daily cross-border movements between European countries. We assume the new strain of the virus appears for the first time in the UK. We simulate the effects of unilateral or multilateral travel bans on the infection curves, focusing on the time until the infection curve peaks and on the peak height. Due to cross-border mobility of people, a more contagious variant of SARS-CoV-2 inevitably spreads across European countries. The variant leads to a peak in the infection curves of all countries observed after about 6 months. The size of observed cross-border flows has a moderate influence on the time until the curve peaks, and has negligible effect on the peak height. We  consider multilateral bans that consist of isolating/quarantining the country where the new strain of the virus has been detected for the first time -- i.e. all European countries imposing travel bans to and from the UK. Such bans slow the spread by increasing the time until the curve peaks by 17 to 18 days. However, they fail to stop the propagation as the virus is already circulating in regions not included in the travel bans. Travel bans hardly impact the height of the peak, which is overwhelmingly determined by local sanitary policies. What is even more remarkable is that the local peak is postponed by 5 additional days only if any country decides to completely self-isolate itself from the rest of the world. By contrast, a unilateral ban -- one EU member alone decides to isolate/quarantine the UK -- has almost no effect on the dynamics of infection as the virus transits through other European countries. In the race against the variants, we conclude that reducing cross-border mobility can buy limited time, helping countries implement better sanitary measures or refine vaccines. This outcome can only be achieved with drastic travel bans and has little influence on the height of the infection curve. In other words, while inducing large economic costs, shutting cross-border mobility has limited effects once patient zero has entered the country. Local sanitary measures matter more: if movers adopt the sanitary measures of the destination country, the shape of the infection curve is mainly determined by local sanitary measures and transmission rates.


Our paper speaks to the recent literature on cross-border mobility and COVID-19 propagation across countries.\footnote{It is worth noting that debates on the role of international migration in the context of an epidemic or a pandemic is not specific to COVID-19. For example, migration is also perceived as a factor explaining the spreading of HIV/AIDS within and across countries (e.g. \cite{anarfi1993, decosas1995, hope2001, ateka2001, brummer2002, docquier2014}. However, these results are not transposable to the COVID-19 crisis as propagation modes are radically different.} On the one hand, several studies show that the spread of the pandemics in Europe cannot be controlled if external borders are permeable, at least in the absence of synchronization of sanitary situations \cite{priesemann2021}. Using modeling approach that combines four computational techniques to assess the effectiveness of non-pharmaceutical interventions (NPIs), \cite{haug2020} argue that social distancing and travel restrictions are among the most effective measures to decrease the reproduction number in all methods. \cite{chen2020} shows that government policies enacted during the Chinese Lunar New Year holiday, during which a massive human migration takes place as individuals travel back to their hometowns, helped reduce the spread of the virus. 
On the other hand, others expressed skepticism about the effectiveness of travel restrictions \cite{lee2020}. Some studies suggest such restrictions can only delay or precipitate disease spread \cite{errett2020}, whereas other research suggests negligible effects on the overall number of cases and show that travel bans only contribute to delaying the overall epidemic progression \cite{chinazzi2020}. Using daily data in a large sample of 175 countries, \cite{asiktas2021} estimates the average dynamic effect of a large set of non-pharmaceutical interventions on the incidence of Covid-19. They find that, on top of restrictions on private gatherings and canceling public events, restrictions on internal movement had no additional effect on the propagation of the virus. Back to the Chinese case, \cite{tian2020} finds that travel restrictions in and out of Wuhan were too late to prevent the spread of the virus to other cities, and \cite{kraemer2020} shows that the correlation between human mobility data and the spatial propagation of the virus drops (and becomes even negative) when accounting for behavioral, clinical and stated interventions.

The paper is organized as follows. The next section details our data on cross-border mobility and provides stylized facts on mobility trends during the pandemic. Then Section \textit{Methods} describes our open-country SIR model and its parameterization. Results of counterfactual mobility experiments are discussed in the following section. These results are discussed in the \textit{Discussion} section.

\section*{Data and stylized facts on cross-border mobility in Europe}
\label{Sec:Data}

To quantify the role of cross-border human mobility (or travel bans) in spreading COVID-19 across European countries, we need estimates of the magnitude of daily cross-border flows of people. We focus on the pre-COVID period and proxy daily flows as the sum of three components, following \cite{docquier2021}. First, we collect data on the average number of commuters by region pair from the Eurostat Labor Force Survey (LFS) and for the year 2019. We use data on commuting outflows by NUTS-2 region, and allocate commuters aged 15 to 64 by destination on the basis of contiguity links with other countries.\footnote{When a NUTS region shares several contiguity links, we share the number of commuters proportionately to the level of income per capita in the destination country.} To identify daily movements, we assume that each commuter spend 4 days per week in the place of work, and multiply LFS numbers by $4/7$. Second, monthly data on air passengers can be obtained from Eurostat by airport of arrival. To identify daily movements, we divide the monthly flow of March 2019 by 31 (the choice of March 2019 is made to avoid a period with many holiday makers). There might be an issue of double counting when aggregating commuting and air passenger data but it should not be too severe as commuters mostly travel by car or by train across contiguous regions. Third, we collect estimates of annual immigration flows from \cite{abel2019} and divide them by 360 to proxy the average number of daily newcomers. For each European country, we aggregate daily commuter, air passenger and migrant inflows.

To identify the dyadic structure of cross-border movements, we rely on Facebook data (FB). FB shares with the scientific community the daily count of the unique numbers of FB users with location services enabled that traveled (by air, train, car, etc.) from an origin country to a destination country during a 24h time period. Only the flows with a minimum of 1,000 users are reported in the data in order to minimize re-identification risk. We use 7-day rolling averages of daily flows to limit the effect of this censoring rule and other mismeasurement problems. We are aware that FB data may not be representative of the whole population. However, it is reassuring that the number of FB users is highly correlated with the population size of the residence country,\footnote{The coefficient of correlation equals 0.97. On average, FB users represent 5.6\% of the population, and this fraction is very stable across countries.}, and the dyadic flows of FB users are highly correlated with the LFS data on dyadic commuting flows. FB flows are almost perfectly symmetric (i.e. flows from $i$ to $j$ are almost equal to flows from $j$ to $i$); this is particularly true after using their 7-day rolling average. This is understandable as the actual underlying flows mostly consist of movements of commuters and business travelers, who cross the border twice within a day. To limit measurement errors, we make FB data perfectly symmetric by taking the mean of the two directional flows. For each destination, we rescale FB inflows as of early March 2020 to match our estimates of actual pre-COVID inflows,\footnote{On average, FB numbers are multiplied by 4.} and use FB share by origin to compute the dyadic structure of daily cross-border movements.

\begin{figure}
\centering
\includegraphics[width=0.8\linewidth]{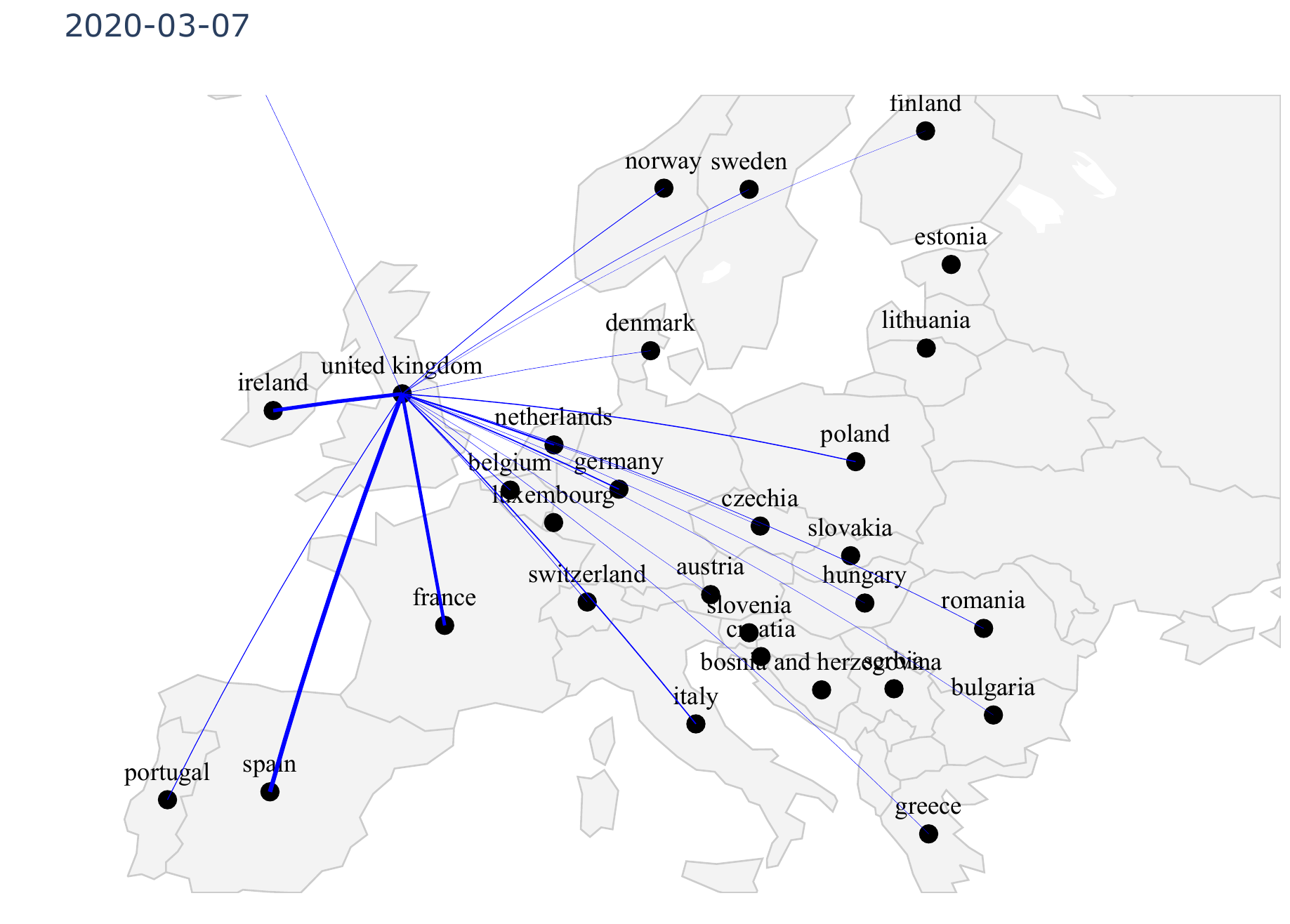}
\caption{Flow map of the average number of daily travelers between the UK and every other country on the first week of March. The width is proportional to the number of flows.}
\label{fig:flowmap_uk}
\end{figure}

Anticipating that our simulation experiments will consist of imposing travel bans between all European countries and the UK, Figure~\ref{fig:flowmap_uk} illustrates the intensity of daily cross-border flows between each European country and the UK.
As can be seen, the UK mainly exchanges its travelers with Ireland, France and Spain. Its connectivity is high enough with all countries in Europe so that a zero patient can originate from the UK. This is the scenario that we observed for the alpha variant.

\begin{figure}
\centering
\includegraphics[width=0.8\linewidth]{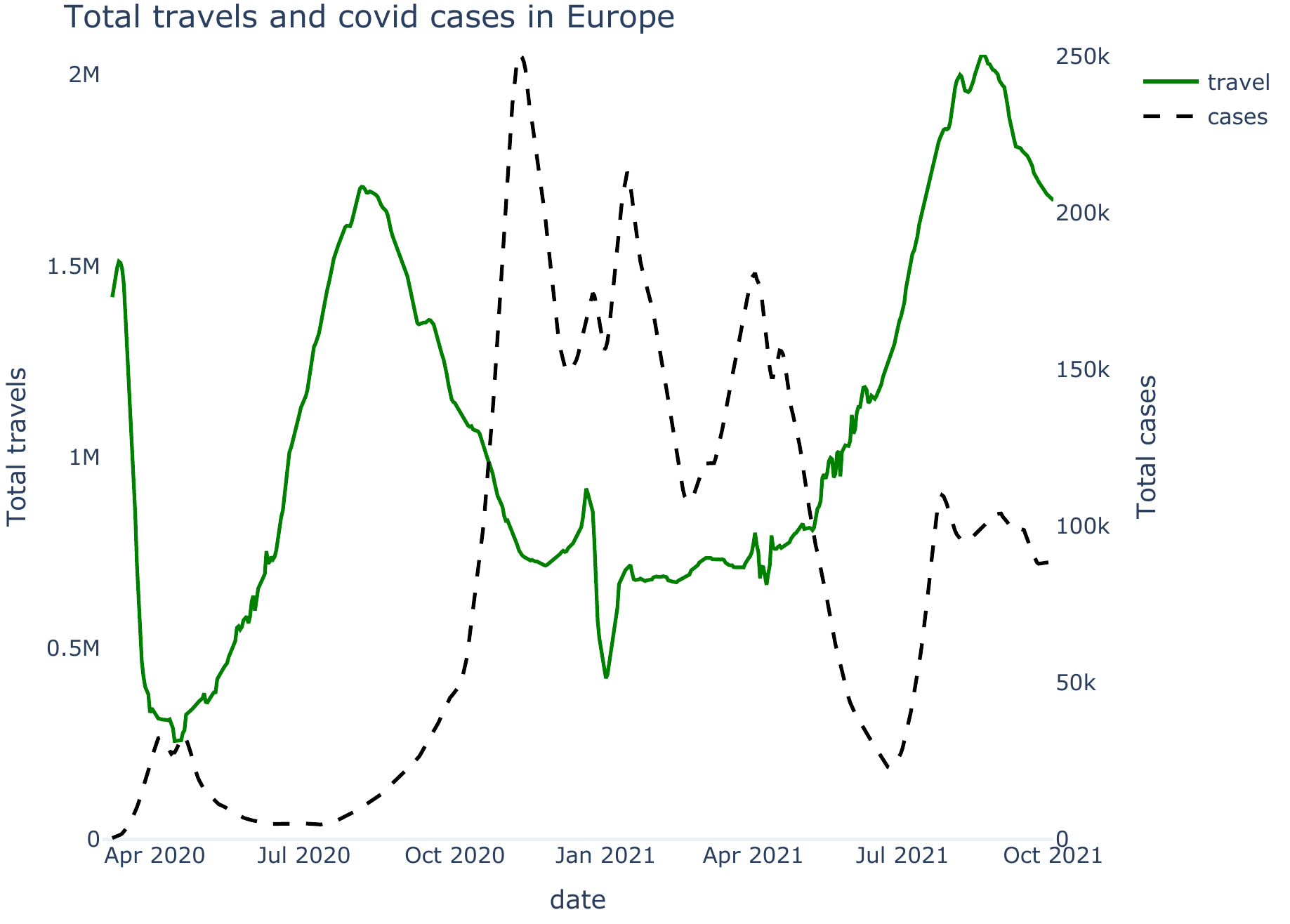}
\caption{Daily evolution of new COVID cases and cross-border mobility in Europe. We identify two points in time: 2020-08-11 (before the start of the second wave) and at 2020-11-22 (at the lowest level of cross-border mobility).}
\label{fig:totravelscases}
\end{figure}

Another advantage of the FB data set is that it documents the daily evolution of cross-border movements during the COVID-19 crisis. Figure~\ref{fig:totravelscases} depicts the daily evolution of (rescaled) dyadic movements observed between the first week of March 2020 and October 2021 (at the very heart of the third wave). Daily movements across European countries declined during the first and second waves of the crisis, and increased during the summer periods. Interestingly, the peak observed in the summer 2021 (when the delta variant spread across countries) exceeds that of August 2021. It is hard to assess whether the decline observed in September 2021 is due to new travel constraints or to the end of the holiday period. Clearly, the correlation between aggregate cross-border flows and the number of new COVID cases is negative, which is very likely to reflect the causal impact of sanitary measures on incentives and permission to travel internationally.

Figure~\ref{fig:totravelscases} illustrates the magnitude if the fall in cross-border mobility associated to each wave of the crisis. This fall reached about 80\% during the first wave (between the first week of March and the first week of May 2020). It was less pronounced during the second wave. When comparing the situation before the second wave (say, on August 11, 2020) and at its peak (say, on November 11, 2020), we identify a 60\% drop in aggregate cross-border movements.



\section*{Methods} 
\label{Sec:Model}

This section describes the model that we used to quantify the effect of travel bans on infection dynamics. As shown in Figure~\ref{fig:totravelscases}, the correlation between daily cross-border movements and new COVID cases is negative. This correlation is the result of bi-directional relationships between daily movements of people and virus propagation: on the one hand, daily movements are endogenous to the sanitary situation governing opportunities to travel internationally; on the other hand, cross-border movements potentially influence the spread of the virus. We care about the latter effect as we aim to quantify the impact of changes in mobility patterns on the international propagation of new COVID-19 variants. 

Using statistical techniques to disentangle the two potentially nonlinear mechanisms at work is difficult. Alternatively, we propose a structural approach and rely on a simple version of the standard compartmental SIR model.\footnote{Using more sophisticated versions of the compartmental model such as the SEIR model of \cite{grimm_extensions_2021} is very unlikely to drastically affect our results.} We develop an open-border, discrete-time SIR model that formalizes the dynamics of the stocks of susceptible ($S_{it}$), infected ($I_{it}$) and recovered ($R_{it}$) individuals across interconnected countries ($i=1,...,N$) over periods of one day ($t=0,...,t_{max}$). Countries are connected through cross-border mobility flows: $\left( m_{ijt}^{S},m_{ijt}^{I},m_{ijt}^{R}\right) $ is a triplet of emigration rates of susceptible, infected and recovered people moving from country $i$ to country $j$ at day $t$. It writes as:

\begin{equation}
\left \{ 
\begin{array}{l}
S_{it+1}=S_{it}\left( 1-\sum_{j\neq i} m_{ijt}^{S}\right) +\sum_{j\neq i}m_{jit}^{S}S_{jt}- \Delta_{it}^{I} \\ 
I_{it+1}=I_{it}\left( 1-\sum_{j\neq i}m_{ijt}^{I}\right) +\sum_{j\neq i}m_{jit}^{I}I_{jt}+\Delta_{it}^{I} - \Delta_{it}^{R} \\ 
R_{it+1}=R_{it+1}\left( 1-\sum_{j\neq i}m_{ijt}^{R}\right) +\sum_{j\neq i}m_{jit}^{R}R_{jt}+ \Delta_{it}^{R}
\end{array}
\right.  \label{eq:sirmob}
\end{equation}
where $\Delta_{it}^{I}$ stands for the flow of new infection in country $i$ at day $t$, and $\Delta_{it}^{R}$ is the daily flow of recovered. Migration flows imply that the population size at time $t+1$ ($P_{it+1}$) can be greater or smaller than the population at time $t$ ($P_{it}$), as it comprises new immigrants from all potential origin countries, $\sum_{j\neq i}\left(m_{jit}^{S}S_{jt}+m_{jit}^{I}I_{jt}+m_{jit}^{R}R_{jt}\right)$, and excludes emigrants to all potential destination countries, $\sum_{j\neq i} \left(m_{ijt}^{S}S_{it} + m_{ijt}^{I}I_{it} + m_{ijt}^{R}R_{it}\right)$. However, by considering symmetric/balanced flows of commuters and business travelers (i.e., inflows and outflows are equal: $m_{ijt}P_{it}=m_{jit}P_{jt}$ $\forall i,j,t$, where $m_{ijt}$ is the average emigration rate to the country $j$ of individuals from the country $i$), we have $P_{it+1}=P_{it}=P_{i}$ in our context. This is important as we aim to express the evolution of the number of active cases and recovered as a percentage of the population.

We model the daily flows of new infections and recovered as following:
\begin{equation}
\left \{ 
\begin{array}{l}
\Delta_{it}^{I}=\frac{\beta _{it}}{P_{it}}\left[ S_{it}\left( 1-\sum_{j}m_{ijt}^{S}\right) +\sum_{j}m_{jit}^{S}S_{jt}\right]
\left[ I_{it}\left( 1-\sum_{j}m_{ijt}^{I}\right) +\sum_{j}m_{jit}^{I}I_{jt}\right] \\ 
\Delta_{it}^{R}= \gamma \left[ I_{it} \left( 1-\sum_{j\neq i}m_{ijt}^{I}\right) + \sum_{j\neq i} m_{jit}^{I}I_{jt} \right]
\end{array}
\right.  \label{eq:flows}
\end{equation}
where $\beta _{it}$ is the daily infection rate, which captures the average number of contacts as well as the probability of transmission in a contact between an infected and a susceptible individuals in country $i$ at day $t$. This probability depends on sanitary measures (NPIs). The factor $\frac{\beta_{i,t}}{P_i}$ is also called the disease transmission rate. The probability of recovery is denoted by $\gamma$ and is assumed to be identical across countries (this parameter is disease-specific). The reproduction number $\rho_{it}$ represents the average number of people that one infected person will pass on a virus to. It can be computed as $\rho_{it} = \frac{\beta_{it}}{\gamma}$.

Modeling daily infections and recoveries as in Eq. (\ref{eq:flows}) implies that movers
at day $t$ contribute to the flows in the destination country, which basically means that they move ``early in the morning'' and fully take part of the destination-country life at day $t$.\footnote{This differs the approach of \cite{liu2013} and \cite{chenmin2020}, who assume that movers at day $t$ contribute to the flows in the origin country, which basically means that they move ``late at night,'' fully take part of the origin-country life at day $t$, and start taking part of the destination-country life at day $t+1$. In the latter framework, we have $\Delta_{it}^{I}=\frac{\beta _{it}S_{it}I_{it}}{P_{it}}$ and $\Delta_{it}^{R}= \gamma I_{it}$.} In the real world, commuters and business travelers share their time between the origin and destination countries.\footnote{In a recent paper, \cite{burzynski2020} develop a multi-sector SIR model in which commuters are exposed to sector-specific infection rates in Luxembourg during working hours, and to origin-specific infection rates when they do not work (family, leisure, social life, etc.). Time allocation depends on employment status and teleworking practices.} The specification choice makes a tiny difference, however, as we use high-frequency (daily) data.

We use numerical experiments to assess how cross-border movements between European countries influence the spread of a new and more contagious variant of the virus. In particular, we care about the effectiveness of potential travel bans between the UK (assumed to be the country where the new strain of the virus appears for the first time) and other European countries. We parameterize our model in a conservative way such that we are likely to overestimate the effect of cross-border movements on the spread of the virus (as well as the effectiveness of travel bans). Our baseline set of assumptions is the following:
\begin{itemize}
    \item The population size of country $i$, $P_i$, is the mean of the levels observed on the 24th of February 2020 and the 12th of February 2021.
    \item We assume that the new variant leads to $\beta_{it}=0.2$ $\forall i,t$ and $\gamma = 0.1$. The parameter $\beta_{it}$ is set rather high and corresponds to an infection rate of COVID-19 in a situation where very limited measures are implemented to reduce the spread of the virus.\footnote{It is important to notice here that this parameter is identical for every country $i$ and every day $t$. It is an unrealistic choice but it allows us to interpret more easily the results of the simulations.} The recovery rate implies that an infected person recovers from the virus in 10 days (close to the actual recovery rate of the COVID-19). This results in a reproduction number of $\rho_{it}=2$.
    \item We assume that the number of recovered/immune person is nil in all countries at time $0$ ($R_{i0}=0$ $\forall i$), and that the number of infected individuals in the UK equals 10 ($I_{UK0}=10$ and $I_{i0}=0$ $\forall i \neq UK$).
    \item Without travel bans, we assume that emigration rates between countries are equal to those observed in the pre-COVID period ($m_{ij0}$ is computed using balanced flows observed between the 29th February 2020 and the 6th of March 2020), as if people would not adjust their mobility behavior to the sanitary conditions.
    \item In practice, infected people are less likely to cross borders as they are more likely to be tested and quarantined, and to spontaneously (self-)withdraw from the labor market if they are sick and symptomatic. Nevertheless, our conservative benchmark trajectory assumes that susceptible, infected and recovered people continue to move with the same intensity ($ m_{ijt}^{S}=m_{ijt}^{I}=m_{ijt}^{R}=m_{ij0}$) $\forall i,j,t$).
\end{itemize}

\begin{figure}
\centering
\includegraphics[width=0.95\linewidth]{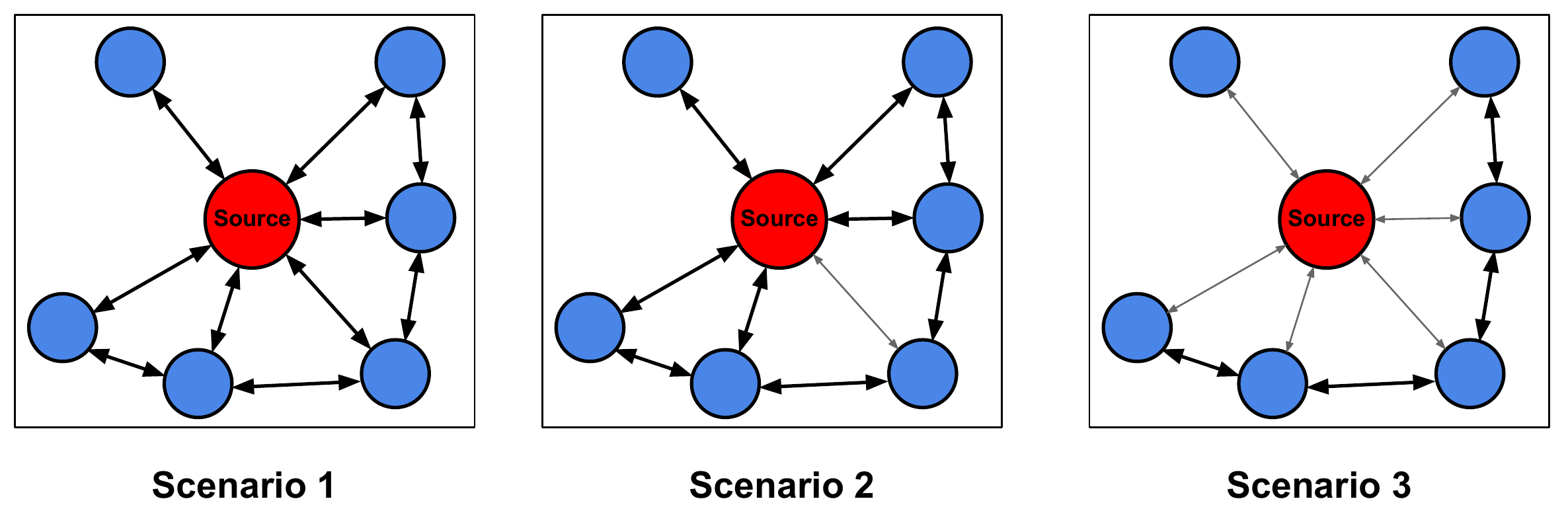}
\caption{Diagrams of the 3 scenarios. The red circle labelled "Source" represents the country that is the origin of the virus or variant (in our case it is the UK). The blue circles represent all the countries sharing daily cross-border flows with the source country.}
\label{fig:scenarios}
\end{figure}

The model is used to simulate the share of infected population in each country during a period of 350 days. The set of assumptions above defines the first baseline (or ``laisser-faire'') scenario (\textbf{Scenario 1} in Figure~\ref{fig:scenarios}). \textbf{Scenario 2} illustrates the situation when a country $i$ alone decides to close its borders with the UK 30 days after the new variant has started to circulate. Given that a complete border closure is unfeasible, we assume that emigration rates from the UK to $i$ and emigration from $i$ to the UK are divided by 10 ($m_{iUKt}=m_{iUK0}/10$ and $m_{UKit}=m_{UKi0}/10$ $\forall t\geq 30$). Scenario 2 corresponds to a unilateral travel ban between the UK and country $i$. Mobility between the UK and the other countries as well as between country $i$ and the other countries is unaffected. \textbf{Scenario 3} considers the case of a multilateral travel ban (i.e., a complete isolation of the UK). All countries jointly decide to close their borders with the UK after 30 days ($m_{iUKt}=m_{iUK0}/10$ and $m_{UKit}=m_{UKi0}/10$ $\forall i$ and $\forall t\geq 30$).

We developed a tool based on this multi-country SIR model available at: \url{https://variantool.herokuapp.com}. Users are allowed to choose a country where a new strain of the virus appears, a common infection rate $\beta_{it}$ and a common recovery rate $\gamma$ for all countries and for the full-time spectrum ($\forall t \in [0,t_{max}]$). Users can also modify the daily migration flows $m_{ijt}$ for any pair ($i,j$) of countries and for any time period $\forall t\in[t_a,t_b]$ with $0\leq t_a \leq t_b \leq t_{max}$. The simulator produces graphs that allow to visualize the evolution of the number of infected people $I_{it}$.

\section*{Results}
\label{Sec:Simul}

The international propagation of COVID-19 variants is obviously linked to cross-border movements. Without cross-border mobility, the modified virus would remain confined in the UK and there would be no patient zero spreading it to other European countries. With cross-border mobility, the propagation of the virus is inevitable. We first illustrate how the new variant spreads over countries when the magnitude of cross-border movements is constant over time and set to the pre-COVID level ($m_{ij0}$). We then consider unilateral and multilateral travel bans and assess their effects on the time until the infection curve peaks and on the peak height.

\begin{figure}[!ht]
\centering
\includegraphics[width=0.95\linewidth]{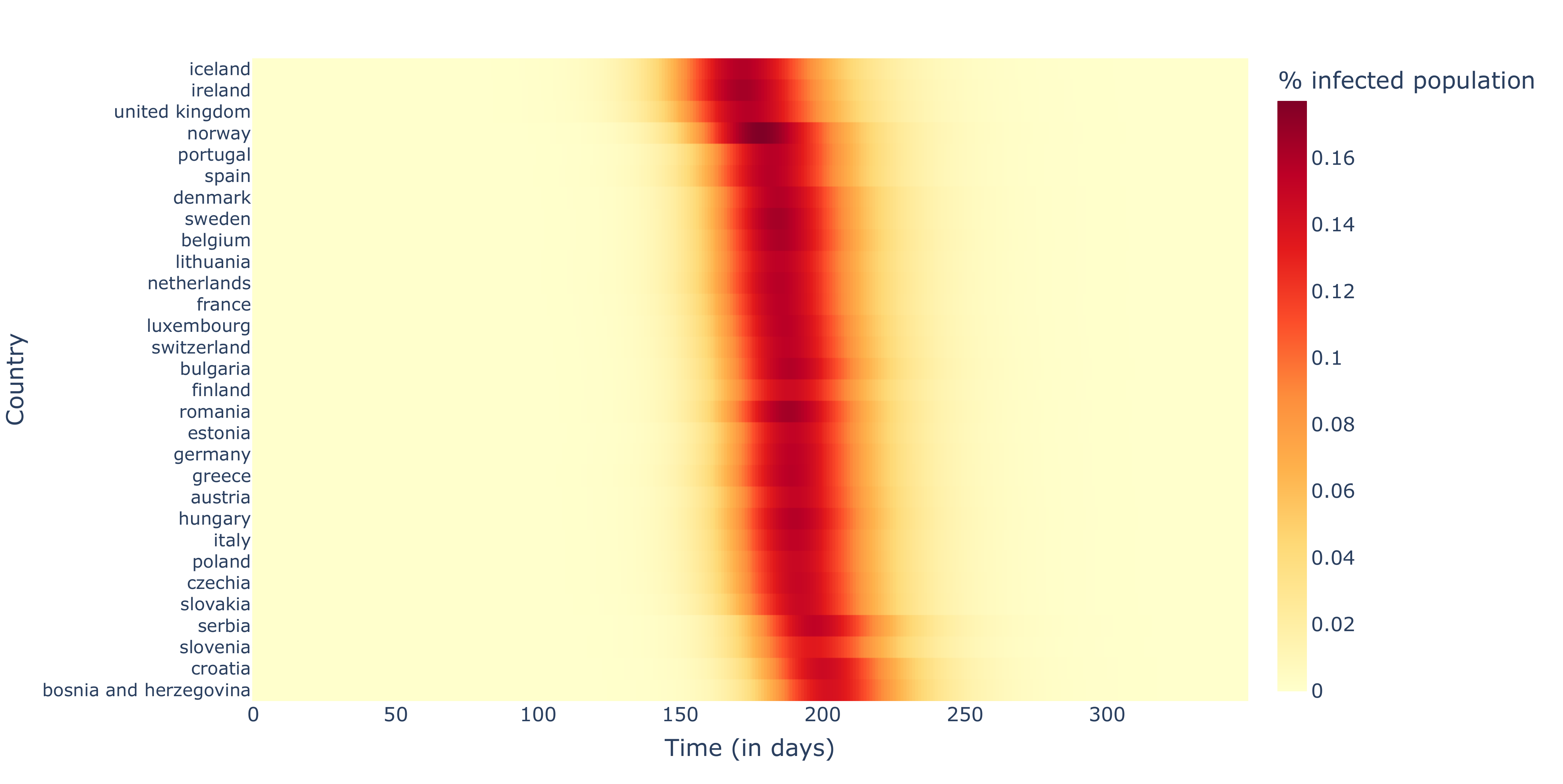}
\includegraphics[width=0.95\linewidth]{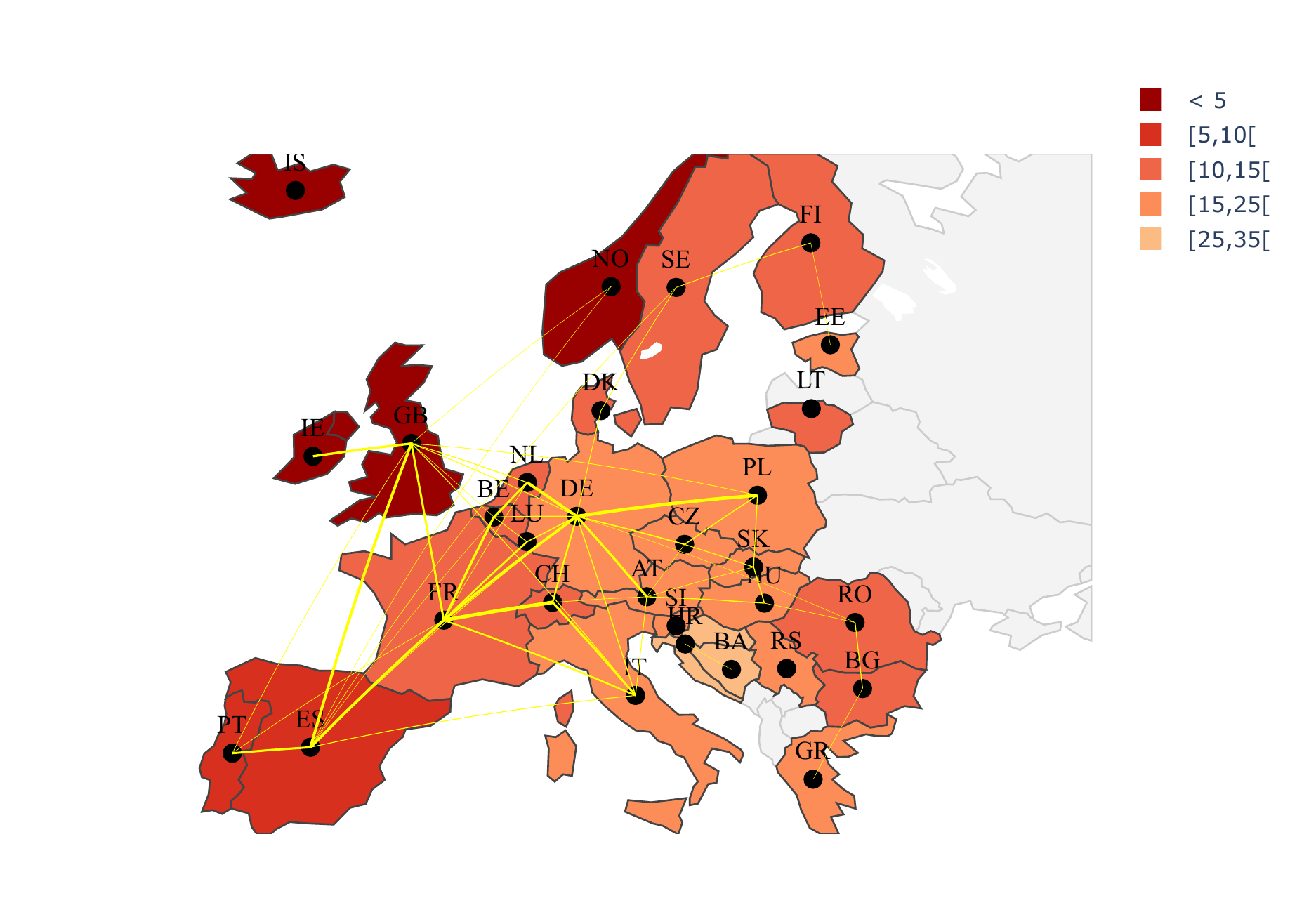}
\caption{Baseline. -- Simulated infection curves in European countries with pre-COVID migration intensity (top panel). Difference with the UK in the number of days until the peak is reached (bottom panel).}
\label{fig:baseline}
\end{figure}

The top panel of figure~\ref{fig:baseline} depicts the infection curve of all European countries over a period of 350 days. We use a heatmap representation: each country corresponds to a row made of 350 daily cells, and the darkness of a cell is a measurement of the share of infected people in the population. With identical/constant transmission rates, all countries experience similar changes in their share of infected people. The peaks of the infection curves are roughly identical (at around 17\%), and the only difference is in the timing of infection. The peak is reached after about 170 days in the UK, where the strain of the virus is first detected.\footnote{Assuming day 0 roughly corresponds to May 1, the UK peak is observed in mid-October.} In the other European countries, the peak can be reached a few days before the UK (say, 5 days in Ireland and Iceland) or up to 30 days later in the case of Bosnia and Herzegovina.

The fact that Ireland and Iceland reach their peak before the UK may seem counter-intuitive. In the case of Ireland, cross-border flows with the UK are large (around 200,000 people cross the border each day) and the population of Ireland is much smaller than that of the UK (roughly 5 million inhabitants against 70 million in the UK). Hence, when patient zero starts infecting susceptible people in Ireland, the share of infected in total population reacts more quickly than in the UK. Although the absolute numbers of infected and recovered people in Ireland are smaller than that of the UK, Ireland will reach its infection peak and herd immunity earlier. The rationale is similar for Iceland. Although the flows of cross-border workers with the UK are 10 times smaller than in Ireland (20,000 daily movers), the population of Iceland is 15 times smaller than the Irish one (340,000 inhabitants).

In the top panel of figure~\ref{fig:baseline}, countries are sorted according to the date of the peak. In the bottom panel, we report the difference with the UK in the number of days until the peak is reached (the darker the country, the smaller the difference), and link it with the magnitude of daily cross-border movements (thickness of the yellow lines). To avoid overloading the map, we only report dyadic flows above 15,000 people. The number of cross-border movers in general, and the ratio of movers to the population in particular, are strong determinants of the timing of infection.


We now turn our attention to unilateral travel bans (Scenario 2). For illustrative purpose, we consider two numerical experiments involving bans between the UK and Belgium or France. First, the flows between the UK and Belgium are divided by 10 at day 30 (i.e. 30 days after the new strain of the virus circulates across countries). The unilateral travel ban has negligible effect on the height of Belgian infection peak and postpones it by one day only, compared with the ``laisser-faire'' path in figure~\ref{fig:baseline}. This indicates that a unilateral travel ban is very poorly effective in slowing the spread of the variant. Belgium is a small country which is strongly connected with many other European countries. After closing the borders with the UK, the variant transits through other European countries, where the infection curve is roughly unaffected.

The map in the bottom of figure~\ref{fig:baseline} shows that France is a central point of European mobility. It exhibits large cross-border flows with many other European countries. Imposing a unilateral travel ban between France and the UK might thus induce larger effects. To represent these bans, the flows between the UK and France are divided by 10 at day 30. The peak of infection in France and in its neighboring countries is postponed by a few days only (3 days in the case of Belgium). This shows that the indirect propagation of the virus can matter more than those induced by bilateral movements (in the case of Belgium), and it confirms that unilateral travel bans are poorly effective, even if implemented by large European countries in the middle of Europe.

\begin{figure}[!ht]
\centering
\includegraphics[width=0.9\linewidth]{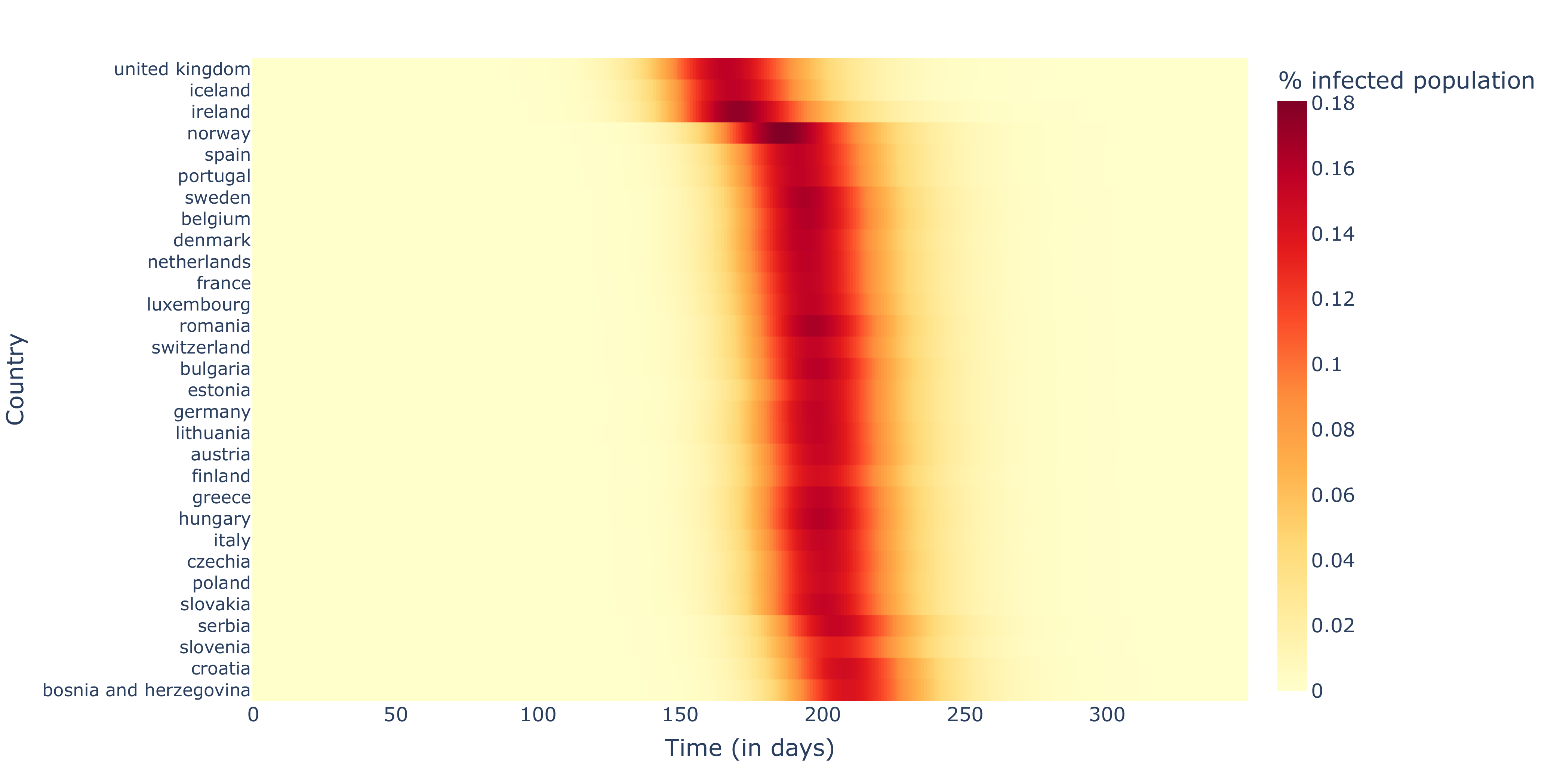}
\includegraphics[width=0.9\linewidth]{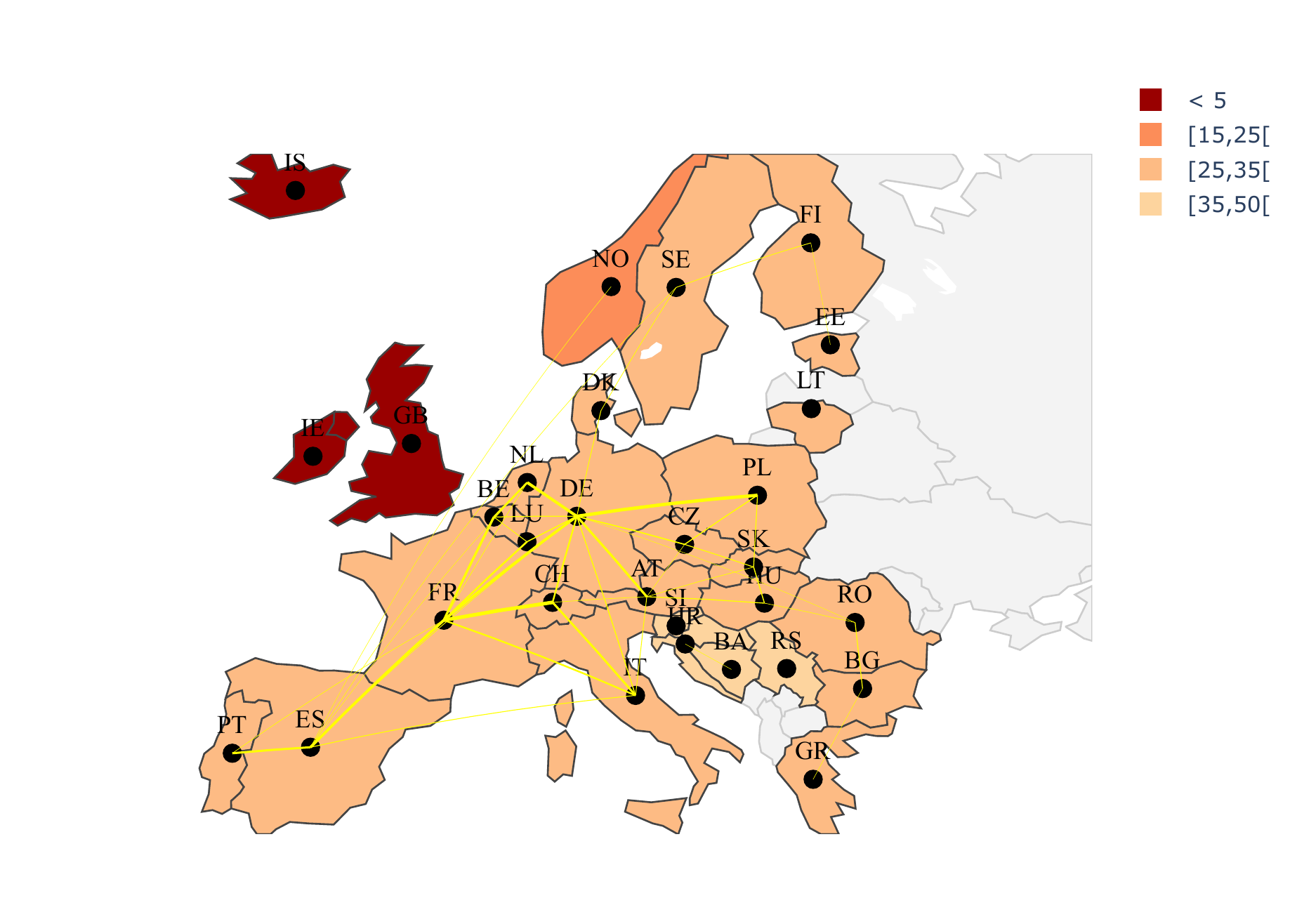}
\caption{Multilateral travel bans. -- Simulated infection curves in European countries when all European countries simultaneously close their borders with the UK after day 30 (top panel). Difference with the UK in the number of days until the peak is reached (bottom panel).}
\label{fig:uk_multilateral}
\end{figure}

In figures~\ref{fig:uk_multilateral}, we simulate the effects of a multilateral travel ban. The flows between the UK and all European countries are divided by 10 at day 30 (Scenario 3). As before, such a multilateral ban hardly influences the height of the peak, at least when movers adopt the sanitary measures of the destination country. The ban only affects the timing of infection. On average, the peak is delayed by 17 to 18 days in European countries. This provides decision-makers with more time to implement sanitary measures or to adapt vaccines. However, multilateral travel bans can buy a limited time, nothing more and nothing less, and this result can only be achieved with drastic restrictions. The benefits appear somewhat limited when compared with the large economic costs that such multilateral bans generate for all borderlands regions in general, and for the target country in particular.

\section*{Discussion}
\label{Sec:Discu}

\begin{figure}[!ht]
\centering
\includegraphics[width=0.95\linewidth]{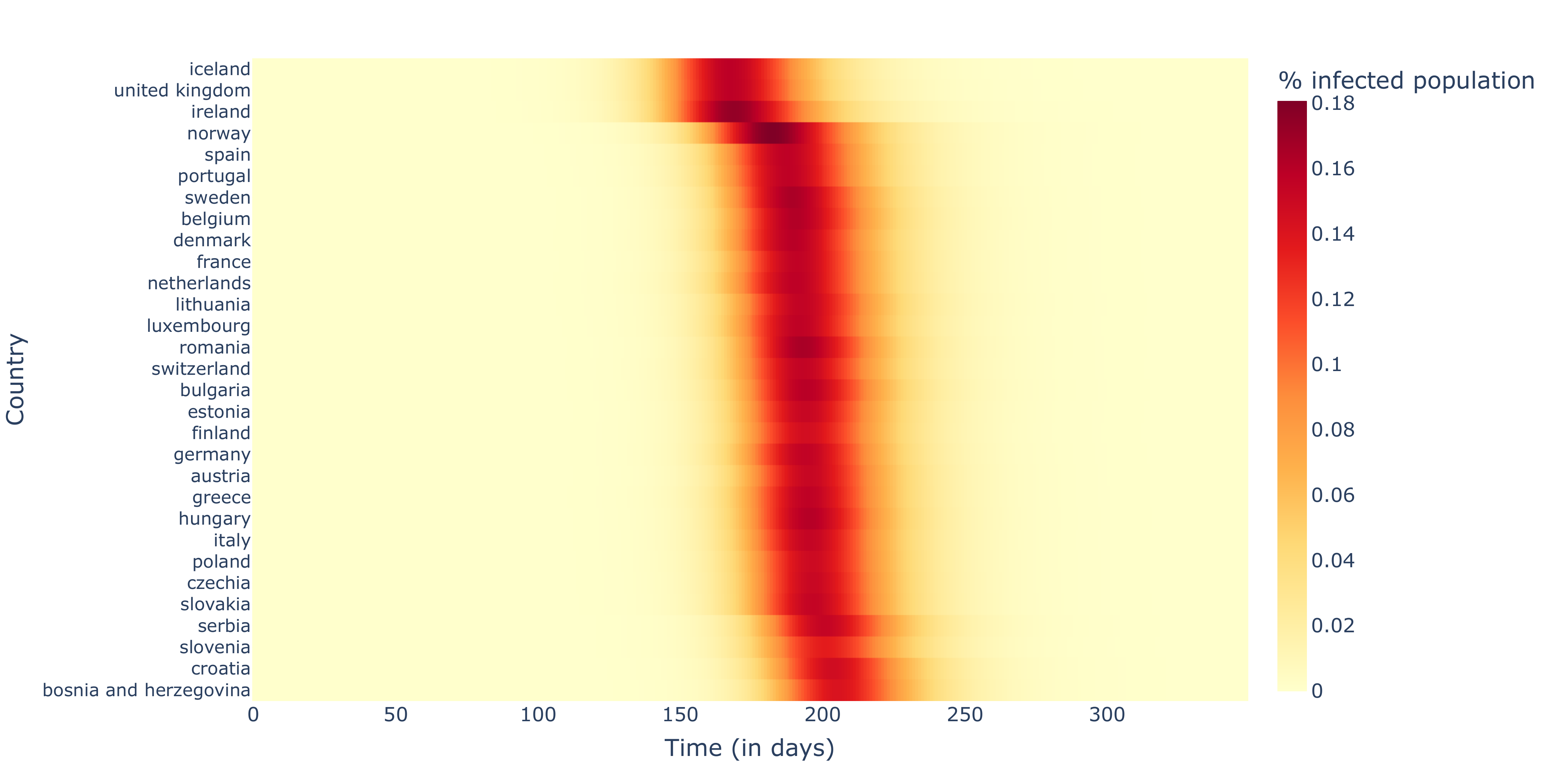}
\includegraphics[width=0.95\linewidth]{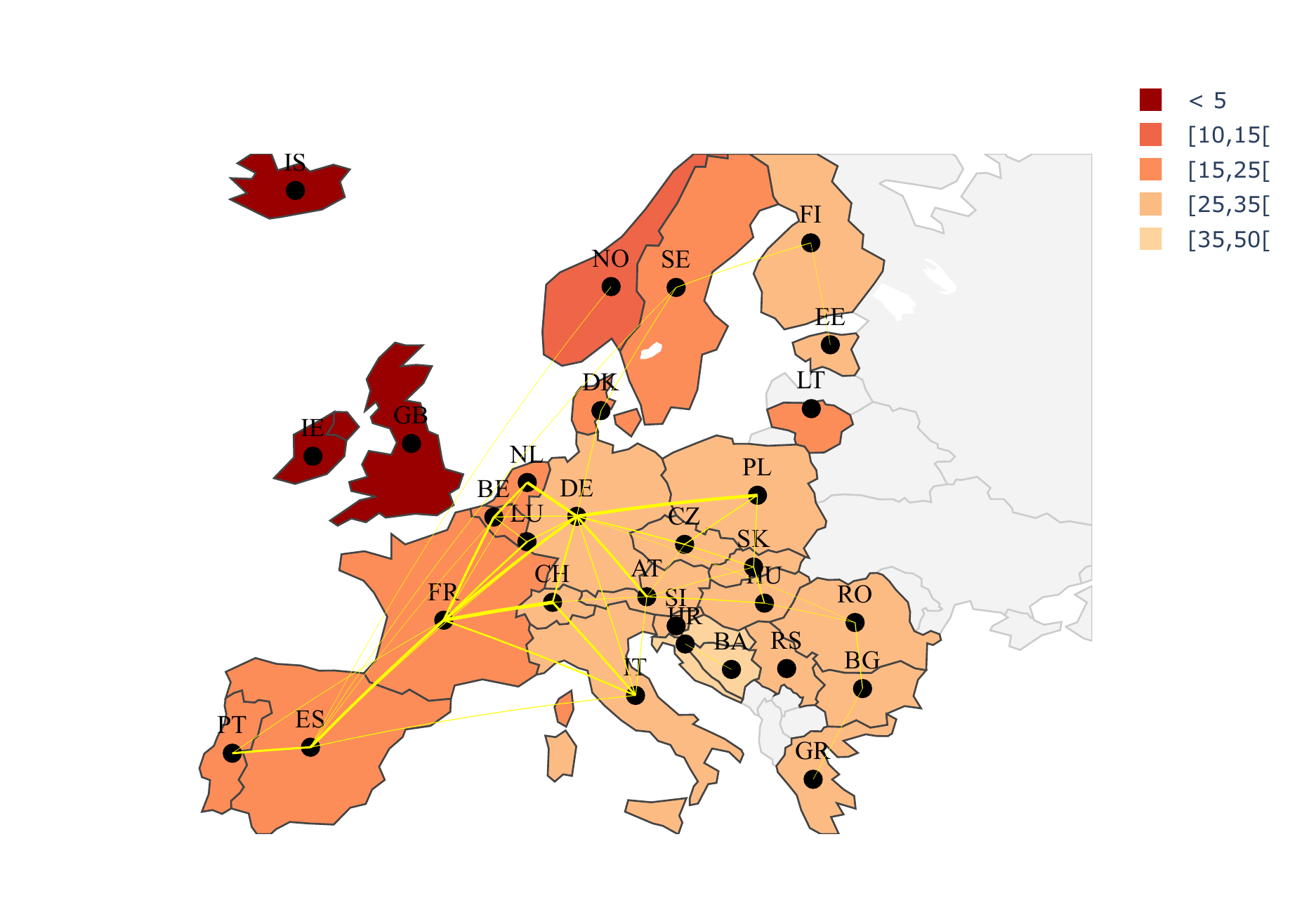}
\caption{Multilateral travel bans. -- Simulated infection curves in European countries when all European countries simultaneously close their borders with the UK after day 60 (top panel). Difference with the UK in the number of days until the peak is reached (bottom panel).}
\label{fig:uk_multilateral60}
\end{figure}

In the race against the variants, reducing cross-border mobility can buy a limited time and help implement better sanitary measures or develop better vaccines. While it seems sensible to limit non-essential movements when epidemiological conditions are deteriorating, our numerical experiments reveal that the effectiveness of such policies is very limited. At best, a generalized travel ban involving both essential and non-essential movements delays the infection peak by two to three weeks. However, achieving this outcome requires a \textbf{fast} and \textbf{drastic} intervention. With regard to the speed of response, somewhat larger effects could be obtained if the policy could be implemented from day 0. Imposing a lag of 30 days between the virus mutation and the travel ban is, however, a very optimistic scenario. WHO data~\cite{noauthor_tracking_nodate} reveal that it took much longer for scientists to detect the past variants of SARS-CoV-2. In Figure~\ref{fig:uk_multilateral60}, we simulate the effect of a multilateral travel ban implemented 60 days after the new variant starts to circulate in the UK and Europe. This delays the peak of 11 to 12 days only, compared with the baseline scenario.\footnote{The longer it takes for European countries to isolate the UK, the shorter the time response. The effect is virtually nil if the travel ban is implemented 150 days after the virus mutation.} With regard to severity of the travel ban, reducing cross-border movements by a factor of 10 corresponds to a very drastic policy.\footnote{The time response to the travel ban would be around 5 days greater if all European countries could totally shut their border after 30 days.} During the first wave of the pandemic (in March-April 2020), the lockdown of European economies translated into a decrease by a factor of 5 (by 3 to 7 depending on the country pair) in aggregate cross-border mobility.

\begin{figure}[!ht]
\centering
\includegraphics[width=0.95\linewidth]{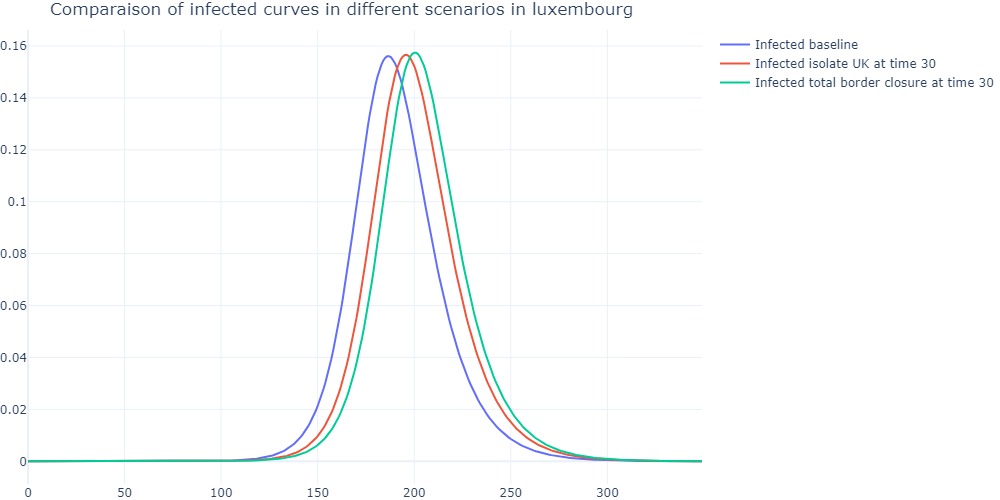}
\includegraphics[width=0.95\linewidth]{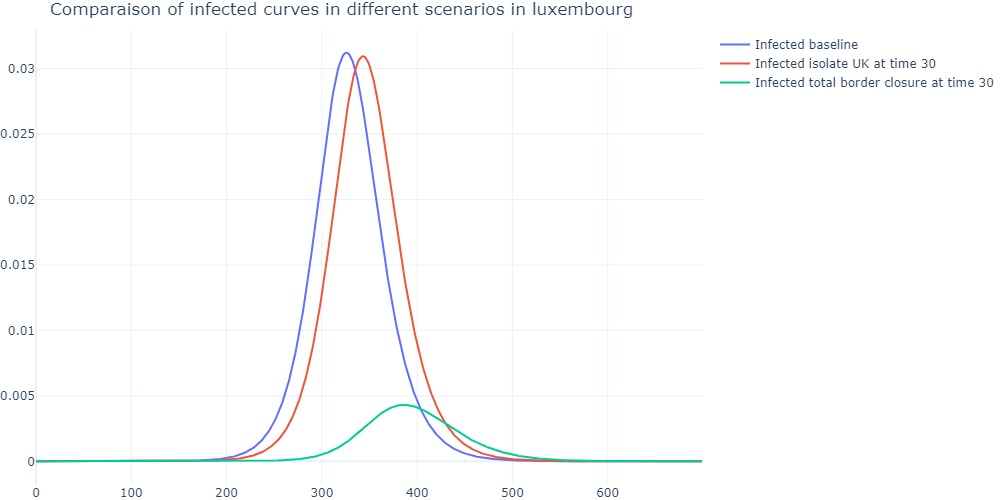}
\caption{Luxembourg case study. -- Infection curves in Luxembourg under three scenarios (baseline, multilateral ban with the UK, multilateral ban with all European countries). The top panel compares the three scenarios without local sanitary responses. The bottom panel compares the three scenarios with $\beta_{LUXt}=0.15$ after day 30.}
\label{fig:luxembourg}
\end{figure}

Another important point for discussion is the difficult tradeoff between the epidemiological benefits from imposing multilateral travel bans and their large economic and welfare losses. Cross-border mobility is vital for many migrants, asylum seekers, commuters or business travelers. It is also a key determinant of economic activity in borderland regions in general, and in national economies in particular. To illustrate this, we focus on the case of Luxembourg, a country where half of the labor force consists of daily commuters and business travelers. Luxembourg hosts 250,000 cross-border movers during each working day. The top panel of figures~\ref{fig:luxembourg} compares the infection curve obtained in the baseline (in blue), when all European countries decide to impose a travel ban with the UK from day 30 (in red), and when Luxembourg unilaterally decides to shut its border with all European countries (in green). Quarantining the UK increases the time until the peak is reached by 17 days. Closing all the borders only delays the peak by 3 to 4 additional days. Preventing this workforce from traveling is likely to induce drastic effects on GDP (a 35 to 40\% loss or so depending on work-from-home practices) and on public finances, if partial unemployment benefits have to be allocated.

Finally, it is important to emphasize that changing mobility patterns has a negligible influence on the height of the peak of the infection curve. Once patient zero has entered the country, the height of the peak is overwhelmingly determined by local sanitary policies (such as mask wearing, hand washing, social distancing, testing and tracing, ventilation of closed spaces... and large-scale vaccination). The bottom panel of figure~\ref{fig:luxembourg} compares the same policies as the top panel, but assumes that Luxembourg decreases its transmission rate ($\beta_{LUXt}$) from 0.20 to 0.15 after day 30. It clearly appears that local sanitary policies are effective in flattening the infection curve, whatever the level of cross-border mobility.

\section*{Conclusion}
\label{Sec:Conclusion}

In response to the spread of the COVID-19 variants, many countries have implemented severe travel bans. Given the large social and economic costs of these measures, we have developed a tool to assess their epidemiological effectiveness in the context of European countries. Our tool is based on a multi-country SIR model that accounts for cross-border flows of people between countries.

Through patient zero, cross-border mobility obviously contributes to propagating new variants across regions and countries. However, given the fact that travel bans are always implemented after the variant has been detected and has circulated for a while, we find that their epidemiological effectiveness is poor, even in countries where cross-border workers and business travelers account for a large share of the labor force. Travel bans hardly impact the height of the peak of the infection curve, which is overwhelmingly determined by local sanitary policies. Unilateral travel bans have negligible effects on the timing and severity of the infection response. Multilateral travel bans can delay the time until the infection curve peaks by a few weeks at best. Although this can help decision-makers implement better sanitary policies, we tend to conclude that travel bans buy limited time while generating huge social and economic costs.


Our work focuses on cross-border movements of people only. On a smaller spatial scale, our results should also apply to within-border movements. Quarantining a city or a region has limited effects on the spread of infection within a country if patient zero has already moved to the other regions. One could argue that within-country mobility is much larger than cross-border mobility. However, we find that travel bans are poorly effective in a country like Luxembourg, where half of the labor force consists of daily commuters and business travelers. Managing new waves of COVID-19 with local sanitary measures applicable to cross-border movers is the most effective option. It induces better epidemiological outcomes and smaller economic cost for all parties concerned.

\newpage


\begin{backmatter}
\section*{Declarations}

\subsection*{Acknowledgements}
The second and third authors acknowledge financial support from the ARC convention on “New approaches to understanding and modeling global migration trends” (convention 18/23-091).

\subsection*{Funding}
Not applicable.

\subsection*{Abbreviations}
EU: European Union \\
FB: Facebook Inc. \\
UK: United Kingdom\\
WHO: World Health Organization\\
IHR: International Health Regulations\\
SIR: Susceptible, Infected, Recovered\\
SEIR: Susceptible, Exposed,  Infected, Recovered\\
NPIs: non-pharmaceutical interventions \\

\subsection*{Availability of data and materials}
The datasets supporting the conclusions of this article are available in the Zenodo repository Multi-Country\_SIR\_Model, \url{https://doi.org/10.5281/zenodo.5783422}

\subsection*{Ethics approval and consent to participate}
No ethics approval or consent required.

\subsection*{Competing interests}
The authors declare that they have no competing interests.

\subsection*{Consent for publication}
Not applicable.

\subsection*{Authors' contributions}
F.D. and P.S. conceived the presented idea. N.G. and P.S. contributed to the design and implementation of the research. F.D. and P.S. collected the data. N.G. cleaned the data, developed the simulation tool and produced the descriptive graphs. All authors contributed to the analysis and discussion of the results, to the writing of the manuscript.


\bibliographystyle{bmc-mathphys} 
\bibliography{TextNico}      


\begin{thebibliography}{23}
\ifx \bisbn   \undefined \def \bisbn  #1{ISBN #1}\fi
\ifx \binits  \undefined \def \binits#1{#1}\fi
\ifx \bauthor  \undefined \def \bauthor#1{#1}\fi
\ifx \batitle  \undefined \def \batitle#1{#1}\fi
\ifx \bjtitle  \undefined \def \bjtitle#1{#1}\fi
\ifx \bvolume  \undefined \def \bvolume#1{\textbf{#1}}\fi
\ifx \byear  \undefined \def \byear#1{#1}\fi
\ifx \bissue  \undefined \def \bissue#1{#1}\fi
\ifx \bfpage  \undefined \def \bfpage#1{#1}\fi
\ifx \blpage  \undefined \def \blpage #1{#1}\fi
\ifx \burl  \undefined \def \burl#1{\textsf{#1}}\fi
\ifx \doiurl  \undefined \def \doiurl#1{\textsf{#1}}\fi
\ifx \betal  \undefined \def \betal{\textit{et al.}}\fi
\ifx \binstitute  \undefined \def \binstitute#1{#1}\fi
\ifx \binstitutionaled  \undefined \def \binstitutionaled#1{#1}\fi
\ifx \bctitle  \undefined \def \bctitle#1{#1}\fi
\ifx \beditor  \undefined \def \beditor#1{#1}\fi
\ifx \bpublisher  \undefined \def \bpublisher#1{#1}\fi
\ifx \bbtitle  \undefined \def \bbtitle#1{#1}\fi
\ifx \bedition  \undefined \def \bedition#1{#1}\fi
\ifx \bseriesno  \undefined \def \bseriesno#1{#1}\fi
\ifx \blocation  \undefined \def \blocation#1{#1}\fi
\ifx \bsertitle  \undefined \def \bsertitle#1{#1}\fi
\ifx \bsnm \undefined \def \bsnm#1{#1}\fi
\ifx \bsuffix \undefined \def \bsuffix#1{#1}\fi
\ifx \bparticle \undefined \def \bparticle#1{#1}\fi
\ifx \barticle \undefined \def \barticle#1{#1}\fi
\ifx \bconfdate \undefined \def \bconfdate #1{#1}\fi
\ifx \botherref \undefined \def \botherref #1{#1}\fi
\ifx \url \undefined \def \url#1{\textsf{#1}}\fi
\ifx \bchapter \undefined \def \bchapter#1{#1}\fi
\ifx \bbook \undefined \def \bbook#1{#1}\fi
\ifx \bcomment \undefined \def \bcomment#1{#1}\fi
\ifx \oauthor \undefined \def \oauthor#1{#1}\fi
\ifx \citeauthoryear \undefined \def \citeauthoryear#1{#1}\fi
\ifx \endbibitem  \undefined \def \endbibitem {}\fi
\ifx \bconflocation  \undefined \def \bconflocation#1{#1}\fi
\ifx \arxivurl  \undefined \def \arxivurl#1{\textsf{#1}}\fi
\csname PreBibitemsHook\endcsname

\bibitem{adhanom_ghebreyesus_who_nodate}
\begin{botherref}
\oauthor{\bsnm{Adhanom~Ghebreyesus}, \binits{T.}}:
{WHO} {Director}-{General}'s opening remarks at the 8th meeting of the {IHR}
  {Emergency} {Committee} on {COVID}-19 – 14 {July} 2021.
\url{https://www.who.int/director-general/speeches/detail/who-director-general-s-opening-remarks-at-the-8th-meeting-of-the-ihr-emergency-committee-on-covid-19-14-july-2021}
Accessed 2021-08-13
\end{botherref}
\endbibitem

\bibitem{noauthor_tracking_nodate}
\begin{botherref}
Tracking {SARS}-{CoV}-2 variants.
\url{https://www.who.int/emergencies/emergency-health-kits/trauma-emergency-surgery-kit-who-tesk-2019/tracking-SARS-CoV-2-variants}
Accessed 2021-08-13
\end{botherref}
\endbibitem

\bibitem{priesemann2021}
\begin{botherref}
\oauthor{\bsnm{Priesemann}, \binits{V.}},
\oauthor{\bsnm{Brinkmann}, \binits{M.M.}},
\oauthor{\bsnm{Ciesek}, \binits{S.}},
\oauthor{\bsnm{Cuschieri}, \binits{S.}},
\oauthor{\bsnm{Czypionka}, \binits{T.}},
\oauthor{\bsnm{Giordano}, \binits{G.}},
\oauthor{\bsnm{Gurdasani}, \binits{D.}},
\oauthor{\bsnm{Hanson}, \binits{C.}},
\oauthor{\bsnm{Hens}, \binits{N.}},
\oauthor{\bsnm{Iftekhar}, \binits{E.}},
\oauthor{\bsnm{Kelly-Irving}, \binits{M.}},
\oauthor{\bsnm{Klimek}, \binits{P.}},
\oauthor{\bsnm{Kretzschmar}, \binits{M.}},
\oauthor{\bsnm{Peichl}, \binits{A.}},
\oauthor{\bsnm{Perc}, \binits{M.}},
\oauthor{\bsnm{Sannino}, \binits{F.}},
\oauthor{\bsnm{Schernhammer}, \binits{E.}},
\oauthor{\bsnm{Schmidt}, \binits{A.}},
\oauthor{\bsnm{Staines}, \binits{A.}},
\oauthor{\bsnm{Szczurek}, \binits{E.}}:
Calling for pan-european commitment for rapid and sustained reduction in
  sars-cov-2 infections.
Lancet,
--1010160140673620326258
(2021)
\end{botherref}
\endbibitem

\bibitem{chen2020}
\begin{barticle}
\bauthor{\bsnm{Chen}, \binits{S.}},
\bauthor{\bsnm{Yang}, \binits{J.}},
\bauthor{\bsnm{Yang}, \binits{W.}},
\bauthor{\bsnm{Wang}, \binits{C.}},
\bauthor{\bsnm{Bärnighausen}, \binits{T.}}:
\batitle{Covid-19 control in china during mass population movements at new
  year}.
\bjtitle{Lancet}
\bvolume{395},
\bfpage{764}--\blpage{766}
(\byear{2020})
\end{barticle}
\endbibitem

\bibitem{lee2020}
\begin{barticle}
\bauthor{\bsnm{Lee}, \binits{K.}},
\bauthor{\bsnm{Worsnop}, \binits{C.Z.}},
\bauthor{\bsnm{Grépin}, \binits{K.A.}},
\bauthor{\bsnm{Kamradt-Scott}, \binits{A.}}:
\batitle{A global coordination on cross-border travel and trade measures
  crucial to covid-19 response}.
\bjtitle{Lancet}
\bvolume{395},
\bfpage{1593}--\blpage{1595}
(\byear{2020})
\end{barticle}
\endbibitem

\bibitem{errett2020}
\begin{barticle}
\bauthor{\bsnm{Errett}, \binits{N.A.}},
\bauthor{\bsnm{Sauer}, \binits{L.M.}},
\bauthor{\bsnm{Rutkow}, \binits{L.}}:
\batitle{An integrative review of the limited evidence of international travel
  bans as an emerging infectious disease disaster control measure}.
\bjtitle{Journal of Emergency Management}
\bvolume{18},
\bfpage{7}--\blpage{14}
(\byear{2020})
\end{barticle}
\endbibitem

\bibitem{chinazzi2020}
\begin{barticle}
\bauthor{\bsnm{Chinazzi}, \binits{M.}},
\bauthor{\bsnm{Davis}, \binits{J.T.}},
\bauthor{\bsnm{Ajelli}, \binits{M.}},
\bauthor{\bsnm{Gioannini}, \binits{C.}},
\bauthor{\bsnm{Litvinova}, \binits{M.}},
\bauthor{\bsnm{Merler}, \binits{S.}},
\bauthor{\bsnm{Piontti}, \binits{A.P.Y.}},
\bauthor{\bsnm{Mu}, \binits{K.}},
\bauthor{\bsnm{Ross}, \binits{L.}},
\bauthor{\bsnm{Sun}, \binits{K.}},
\bauthor{\bsnm{Viboud}, \binits{C.}},
\bauthor{\bsnm{Xiong}, \binits{X.}},
\bauthor{\bsnm{Yu}, \binits{H.}},
\bauthor{\bsnm{Halloran}, \binits{M.E.}},
\bauthor{\bsnm{Longini~Jr}, \binits{I.M.}},
\bauthor{\bsnm{Vespignani}, \binits{A.}}:
\batitle{The effect of travel restrictions on the spread of the 2019 novel
  coronavirus (covid-19) outbreak}.
\bjtitle{Science}
\bvolume{368},
\bfpage{395}--\blpage{400}
(\byear{2020})
\end{barticle}
\endbibitem

\bibitem{asiktas2021}
\begin{barticle}
\bauthor{\bsnm{Askitas}, \binits{N.}},
\bauthor{\bsnm{Tatsiramos}, \binits{K.}},
\bauthor{\bsnm{Verheyden}, \binits{B.}}:
\batitle{Estimating worldwide effects of non‐pharmaceutical interventions on
  covid‐19 incidence and population mobility patterns using a
  multiple‐event study}.
\bjtitle{Scientific Reports}
\bvolume{11}(\bissue{1972}),
\bfpage{41598}--\blpage{02181442}
(\byear{2021})
\end{barticle}
\endbibitem

\bibitem{tian2020}
\begin{barticle}
\bauthor{\bsnm{Tian}, \binits{H.}},
\bauthor{\bsnm{Liu}, \binits{Y.}},
\bauthor{\bsnm{Li}, \binits{Y.}},
\bauthor{\bsnm{Wu2}, \binits{C.-H.}},
\bauthor{\bsnm{Chen}, \binits{B.}},
\bauthor{\bsnm{Kraemer}, \binits{M.U.G.}},
\bauthor{\bsnm{Li}, \binits{B.}},
\bauthor{\bsnm{Cai}, \binits{J.}},
\bauthor{\bsnm{Xu}, \binits{B.}},
\bauthor{\bsnm{Yang}, \binits{Q.}},
\bauthor{\bsnm{Wang}, \binits{B.}},
\bauthor{\bsnm{Yang}, \binits{P.}},
\bauthor{\bsnm{Cui}, \binits{Y.}},
\bauthor{\bsnm{Song}, \binits{Y.}},
\bauthor{\bsnm{Zheng}, \binits{P.}},
\bauthor{\bsnm{Wang}, \binits{Q.}},
\bauthor{\bsnm{Bjornstad}, \binits{O.N.}},
\bauthor{\bsnm{Yang}, \binits{R.}},
\bauthor{\bsnm{Grenfell}, \binits{B.T.}},
\bauthor{\bsnm{Pybus}, \binits{O.G.}},
\bauthor{\bsnm{Dye}, \binits{C.}}:
\batitle{An investigation of transmission control measures during the first 50
  days of the covid-19 epidemic in china}.
\bjtitle{Science}
\bvolume{368},
\bfpage{638}--\blpage{642}
(\byear{2020})
\end{barticle}
\endbibitem

\bibitem{kraemer2020}
\begin{barticle}
\bauthor{\bsnm{Kraemer}, \binits{M.U.G.}},
\bauthor{\bsnm{Yang}, \binits{C.-H.}},
\bauthor{\bsnm{Gutierrez}, \binits{B.}},
\bauthor{\bsnm{Wu}, \binits{C.-H.}},
\bauthor{\bsnm{Klein}, \binits{B.}},
\bauthor{\bsnm{Pigott}, \binits{D.M.}},
\bauthor{\bsnm{Group}, \binits{O.C.-.D.W.}},
\bauthor{\bparticle{du} \bsnm{Plessis}, \binits{L.}},
\bauthor{\bsnm{Faria}, \binits{N.R.}},
\bauthor{\bsnm{Li}, \binits{R.}},
\bauthor{\bsnm{Hanage}, \binits{W.P.}},
\bauthor{\bsnm{Brownstein}, \binits{J.S.}},
\bauthor{\bsnm{Layan}, \binits{M.}},
\bauthor{\bsnm{Vespignani}, \binits{A.}},
\bauthor{\bsnm{Tian}, \binits{H.}},
\bauthor{\bsnm{Dye}, \binits{C.}},
\bauthor{\bsnm{Pybus}, \binits{O.G.}},
\bauthor{\bsnm{Scarpino}, \binits{S.V.}}:
\batitle{The effect of human mobility and control measures on the covid-19
  epidemic in china}.
\bjtitle{Science}
\bvolume{497},
\bfpage{493}--\blpage{497}
(\byear{2020})
\end{barticle}
\endbibitem

\bibitem{anarfi1993}
\begin{barticle}
\bauthor{\bsnm{Anarfi}, \binits{J.K.}}:
\batitle{Sexuality, migration and aids in ghana: a socio-behavioural study}.
\bjtitle{Health Transition Review}
\bvolume{3},
\bfpage{1}--\blpage{22}
(\byear{1993})
\end{barticle}
\endbibitem

\bibitem{decosas1995}
\begin{barticle}
\bauthor{\bsnm{Decosas}, \binits{J.}},
\bauthor{\bsnm{Kane}, \binits{F.}},
\bauthor{\bsnm{Anarfi}, \binits{J.K.}},
\bauthor{\bsnm{Sodji}, \binits{K.D.}},
\bauthor{\bsnm{Wagner}, \binits{H.U.}}:
\batitle{Migration and aids}.
\bjtitle{Lancet}
\bvolume{346}(\bissue{8978}),
\bfpage{826}--\blpage{828}
(\byear{1995})
\end{barticle}
\endbibitem

\bibitem{hope2001}
\begin{barticle}
\bauthor{\bsnm{Hope}, \binits{K.R.}}:
\batitle{Population mobility and multi-partner sex in botswana: implications
  for a spread of hiv/aids}.
\bjtitle{Journal of Reproductive Health}
\bvolume{5}(\bissue{3}),
\bfpage{73}--\blpage{93}
(\byear{2001})
\end{barticle}
\endbibitem

\bibitem{ateka2001}
\begin{barticle}
\bauthor{\bsnm{Ateka}, \binits{G.K.}}:
\batitle{Factors in hiv/aids transmission in sub-saharan africa}.
\bjtitle{Bulletin of the World Health Organization}
\bvolume{79}(\bissue{12}),
\bfpage{1168}
(\byear{2001})
\end{barticle}
\endbibitem

\bibitem{brummer2002}
\begin{botherref}
\oauthor{\bsnm{Brummer}, \binits{D.}}:
Labour migration and hiv/aids in southern africa.
IOM’s Regional Office for Southern Africa: Geneva
(2002)
\end{botherref}
\endbibitem

\bibitem{docquier2014}
\begin{barticle}
\bauthor{\bsnm{Docquier}, \binits{F.}},
\bauthor{\bsnm{Vasilakis}, \binits{C.}},
\bauthor{\bsnm{Tamfutu~Munsi}, \binits{D.}}:
\batitle{International migration and the propagation of hiv in sub-saharan
  africa}.
\bjtitle{Journal of Health Economics}
\bvolume{35}(\bissue{8978}),
\bfpage{20}--\blpage{34}
(\byear{2014})
\end{barticle}
\endbibitem

\bibitem{haug2020}
\begin{barticle}
\bauthor{\bsnm{Haug}, \binits{N.}},
\bauthor{\bsnm{Geyrhofer}, \binits{L.}},
\bauthor{\bsnm{Londei}, \binits{A.}},
\bauthor{\bsnm{Dervic}, \binits{E.}},
\bauthor{\bsnm{Desvars-Larrive}, \binits{A.}},
\bauthor{\bsnm{Loreto}, \binits{V.}},
\bauthor{\bsnm{Beate~Pinior}, \binits{B.} \bsuffix{Thurner}},
\bauthor{\bsnm{Stefan}},
\bauthor{\bsnm{Klimek}, \binits{P.}}:
\batitle{Ranking the effectiveness of worldwide covid-19 government
  interventions}.
\bjtitle{Nature Human Behaviour}
\bvolume{4},
\bfpage{1305}--\blpage{1312}
(\byear{2020})
\end{barticle}
\endbibitem

\bibitem{docquier2021}
\begin{botherref}
\oauthor{\bsnm{Docquier}, \binits{F.}},
\oauthor{\bsnm{Golenvaux}, \binits{N.}},
\oauthor{\bsnm{Nijssen}, \binits{S.}},
\oauthor{\bsnm{Schau}, \binits{P.}},
\oauthor{\bsnm{Stips}, \binits{F.}}:
Cross-border mobility responses to covid-19 in europe: New evidence from
  facebook data.
Manuscript (Universit\'{e} catholique de Louvain)
(2021)
\end{botherref}
\endbibitem

\bibitem{abel2019}
\begin{barticle}
\bauthor{\bsnm{Abel}, \binits{G.J.}},
\bauthor{\bsnm{Cohen}, \binits{J.E.}}:
\batitle{Bilateral international migration flow estimates for 200 countries}.
\bjtitle{Scientific data}
\bvolume{6}(\bissue{1}),
\bfpage{1}--\blpage{13}
(\byear{2019})
\end{barticle}
\endbibitem

\bibitem{grimm_extensions_2021}
\begin{barticle}
\bauthor{\bsnm{Grimm}, \binits{V.}},
\bauthor{\bsnm{Mengel}, \binits{F.}},
\bauthor{\bsnm{Schmidt}, \binits{M.}}:
\batitle{Extensions of the {SEIR} model for the analysis of tailored social
  distancing and tracing approaches to cope with {COVID}-19}.
\bjtitle{Scientific Reports}
\bvolume{11}(\bissue{1}),
\bfpage{4214}
(\byear{2021}).
doi:\doiurl{10.1038/s41598-021-83540-2}.
\bcomment{Bandiera\_abtest: a Cc\_license\_type: cc\_by Cg\_type: Nature
  Research Journals Number: 1 Primary\_atype: Research Publisher: Nature
  Publishing Group Subject\_term: Epidemiology;Experimental evolution
  Subject\_term\_id: epidemiology;experimental-evolution}.
Accessed 2021-10-04
\end{barticle}
\endbibitem

\bibitem{liu2013}
\begin{barticle}
\bauthor{\bsnm{Liu}, \binits{M.}},
\bauthor{\bsnm{Xiao}, \binits{Y.}}:
\batitle{Modeling and analysis of epidemic diffusion with population
  migration}.
\bjtitle{Journal of Applied Mathematics}
\bvolume{583648},
\bfpage{8}
(\byear{2013})
\end{barticle}
\endbibitem

\bibitem{chenmin2020}
\begin{barticle}
\bauthor{\bsnm{Chen}, \binits{M.}},
\bauthor{\bsnm{Miao}, \binits{L.}},
\bauthor{\bsnm{Hao}, \binits{Y.}},
\bauthor{\bsnm{Liu}, \binits{Z.}},
\bauthor{\bsnm{Hu}, \binits{L.}},
\bauthor{\bsnm{Wang}, \binits{L.}}:
\batitle{The introduction of population migration to seiar for covid-19
  epidemic modeling with an efficient intervention strategy}.
\bjtitle{Information Fusion}
\bvolume{64},
\bfpage{252}--\blpage{258}
(\byear{2020})
\end{barticle}
\endbibitem

\bibitem{burzynski2020}
\begin{barticle}
\bauthor{\bsnm{Burzynski}, \binits{M.}},
\bauthor{\bsnm{Machado}, \binits{J.}},
\bauthor{\bsnm{Aalto}, \binits{A.}},
\bauthor{\bsnm{Beine}, \binits{M.}},
\bauthor{\bsnm{Haas}, \binits{T.}},
\bauthor{\bsnm{Kemp}, \binits{F.}},
\bauthor{\bsnm{Magni}, \binits{S.}},
\bauthor{\bsnm{Mombaerts}, \binits{L.}},
\bauthor{\bsnm{Picard}, \binits{P.}},
\bauthor{\bsnm{Proverbio}, \binits{D.}},
\bauthor{\bsnm{Skupin}, \binits{A.}},
\bauthor{\bsnm{Docquier}, \binits{F.}}:
\batitle{Covid-19 crisis management in luxembourg: Insights from an
  epidemionomic approach}.
\bjtitle{Economics and Human Biology}
\bvolume{43},
\bfpage{101051}
(\byear{2021})
\end{barticle}
\endbibitem

\end{thebibliography}

\newcommand{\BMCxmlcomment}[1]{}

\BMCxmlcomment{

<refgrp>

<bibl id="B1">
  <title><p>{WHO} {Director}-{General}'s opening remarks at the 8th meeting of
  the {IHR} {Emergency} {Committee} on {COVID}-19 – 14 {July}
  2021</p></title>
  <aug>
    <au><snm>Adhanom Ghebreyesus</snm><fnm>T</fnm></au>
  </aug>
  <url>https://www.who.int/director-general/speeches/detail/who-director-general-s-opening-remarks-at-the-8th-meeting-of-the-ihr-emergency-committee-on-covid-19-14-july-2021</url>
</bibl>

<bibl id="B2">
  <title><p>Tracking {SARS}-{CoV}-2 variants</p></title>
  <url>https://www.who.int/emergencies/emergency-health-kits/trauma-emergency-surgery-kit-who-tesk-2019/tracking-SARS-CoV-2-variants</url>
</bibl>

<bibl id="B3">
  <title><p>Calling for pan-European commitment for rapid and sustained
  reduction in SARS-CoV-2 infections</p></title>
  <aug>
    <au><snm>Priesemann</snm><fnm>V</fnm></au>
    <au><snm>Brinkmann</snm><fnm>MM</fnm></au>
    <au><snm>Ciesek</snm><fnm>S</fnm></au>
    <au><snm>Cuschieri</snm><fnm>S</fnm></au>
    <au><snm>Czypionka</snm><fnm>T</fnm></au>
    <au><snm>Giordano</snm><fnm>G</fnm></au>
    <au><snm>Gurdasani</snm><fnm>D</fnm></au>
    <au><snm>Hanson</snm><fnm>C</fnm></au>
    <au><snm>Hens</snm><fnm>N</fnm></au>
    <au><snm>Iftekhar</snm><fnm>E</fnm></au>
    <au><snm>Kelly Irving</snm><fnm>M</fnm></au>
    <au><snm>Klimek</snm><fnm>P</fnm></au>
    <au><snm>Kretzschmar</snm><fnm>M</fnm></au>
    <au><snm>Peichl</snm><fnm>A</fnm></au>
    <au><snm>Perc</snm><fnm>M</fnm></au>
    <au><snm>Sannino</snm><fnm>F</fnm></au>
    <au><snm>Schernhammer</snm><fnm>E</fnm></au>
    <au><snm>Schmidt</snm><fnm>A</fnm></au>
    <au><snm>Staines</snm><fnm>A</fnm></au>
    <au><snm>Szczurek</snm><fnm>E</fnm></au>
  </aug>
  <source>Lancet</source>
  <pubdate>2021</pubdate>
  <fpage>forthcoming,https://doi.org/10.1016/S0140</fpage>
  <lpage>6736(20)32625-8</lpage>
</bibl>

<bibl id="B4">
  <title><p>COVID-19 control in China during mass population movements at New
  Year</p></title>
  <aug>
    <au><snm>Chen</snm><fnm>S</fnm></au>
    <au><snm>Yang</snm><fnm>J</fnm></au>
    <au><snm>Yang</snm><fnm>W</fnm></au>
    <au><snm>Wang</snm><fnm>C</fnm></au>
    <au><snm>Bärnighausen</snm><fnm>T</fnm></au>
  </aug>
  <source>Lancet</source>
  <pubdate>2020</pubdate>
  <volume>395</volume>
  <fpage>764</fpage>
  <lpage>-766</lpage>
</bibl>

<bibl id="B5">
  <title><p>A Global coordination on cross-border travel and trade measures
  crucial to COVID-19 response</p></title>
  <aug>
    <au><snm>Lee</snm><fnm>K</fnm></au>
    <au><snm>Worsnop</snm><fnm>CZ</fnm></au>
    <au><snm>Grépin</snm><fnm>KA</fnm></au>
    <au><snm>Kamradt Scott</snm><fnm>A</fnm></au>
  </aug>
  <source>Lancet</source>
  <pubdate>2020</pubdate>
  <volume>395</volume>
  <fpage>1593</fpage>
  <lpage>-1595</lpage>
</bibl>

<bibl id="B6">
  <title><p>An integrative review of the limited evidence of international
  travel bans as an emerging infectious disease disaster control
  measure</p></title>
  <aug>
    <au><snm>Errett</snm><fnm>NA</fnm></au>
    <au><snm>Sauer</snm><fnm>LM</fnm></au>
    <au><snm>Rutkow</snm><fnm>L</fnm></au>
  </aug>
  <source>Journal of Emergency Management</source>
  <pubdate>2020</pubdate>
  <volume>18</volume>
  <fpage>7</fpage>
  <lpage>-14</lpage>
</bibl>

<bibl id="B7">
  <title><p>The effect of travel restrictions on the spread of the 2019 novel
  coronavirus (COVID-19) outbreak</p></title>
  <aug>
    <au><snm>Chinazzi</snm><fnm>M</fnm></au>
    <au><snm>Davis</snm><fnm>JT</fnm></au>
    <au><snm>Ajelli</snm><fnm>M</fnm></au>
    <au><snm>Gioannini</snm><fnm>C</fnm></au>
    <au><snm>Litvinova</snm><fnm>M</fnm></au>
    <au><snm>Merler</snm><fnm>S</fnm></au>
    <au><snm>Piontti</snm><fnm>APY</fnm></au>
    <au><snm>Mu</snm><fnm>K</fnm></au>
    <au><snm>Ross</snm><fnm>L</fnm></au>
    <au><snm>Sun</snm><fnm>K</fnm></au>
    <au><snm>Viboud</snm><fnm>C</fnm></au>
    <au><snm>Xiong</snm><fnm>X</fnm></au>
    <au><snm>Yu</snm><fnm>H</fnm></au>
    <au><snm>Halloran</snm><fnm>ME</fnm></au>
    <au><snm>Longini Jr</snm><fnm>IM</fnm></au>
    <au><snm>Vespignani</snm><fnm>A</fnm></au>
  </aug>
  <source>Science</source>
  <pubdate>2020</pubdate>
  <volume>368</volume>
  <fpage>395</fpage>
  <lpage>-400</lpage>
</bibl>

<bibl id="B8">
  <title><p>Estimating worldwide effects of non‐pharmaceutical interventions
  on COVID‐19 incidence and population mobility patterns using a
  multiple‐event study</p></title>
  <aug>
    <au><snm>Askitas</snm><fnm>N</fnm></au>
    <au><snm>Tatsiramos</snm><fnm>K</fnm></au>
    <au><snm>Verheyden</snm><fnm>B</fnm></au>
  </aug>
  <source>Scientific Reports</source>
  <pubdate>2021</pubdate>
  <volume>11</volume>
  <issue>1972</issue>
  <fpage>https://www.nature.com/articles/s41598</fpage>
  <lpage>021-81442-x</lpage>
</bibl>

<bibl id="B9">
  <title><p>An investigation of transmission control measures during the first
  50 days of the COVID-19 epidemic in China</p></title>
  <aug>
    <au><snm>Tian</snm><fnm>H</fnm></au>
    <au><snm>Liu</snm><fnm>Y</fnm></au>
    <au><snm>Li</snm><fnm>Y</fnm></au>
    <au><snm>Wu2</snm><fnm>CH</fnm></au>
    <au><snm>Chen</snm><fnm>B</fnm></au>
    <au><snm>Kraemer</snm><fnm>MUG</fnm></au>
    <au><snm>Li</snm><fnm>B</fnm></au>
    <au><snm>Cai</snm><fnm>J</fnm></au>
    <au><snm>Xu</snm><fnm>B</fnm></au>
    <au><snm>Yang</snm><fnm>Q</fnm></au>
    <au><snm>Wang</snm><fnm>B</fnm></au>
    <au><snm>Yang</snm><fnm>P</fnm></au>
    <au><snm>Cui</snm><fnm>Y</fnm></au>
    <au><snm>Song</snm><fnm>Y</fnm></au>
    <au><snm>Zheng</snm><fnm>P</fnm></au>
    <au><snm>Wang</snm><fnm>Q</fnm></au>
    <au><snm>Bjornstad</snm><fnm>ON</fnm></au>
    <au><snm>Yang</snm><fnm>R</fnm></au>
    <au><snm>Grenfell</snm><fnm>BT</fnm></au>
    <au><snm>Pybus</snm><fnm>OG</fnm></au>
    <au><snm>Dye</snm><fnm>C</fnm></au>
  </aug>
  <source>Science</source>
  <pubdate>2020</pubdate>
  <volume>368</volume>
  <fpage>638</fpage>
  <lpage>-642</lpage>
</bibl>

<bibl id="B10">
  <title><p>The effect of human mobility and control measures on the COVID-19
  epidemic in China</p></title>
  <aug>
    <au><snm>Kraemer</snm><fnm>MUG</fnm></au>
    <au><snm>Yang</snm><fnm>CH</fnm></au>
    <au><snm>Gutierrez</snm><fnm>B</fnm></au>
    <au><snm>Wu</snm><fnm>CH</fnm></au>
    <au><snm>Klein</snm><fnm>B</fnm></au>
    <au><snm>Pigott</snm><fnm>DM</fnm></au>
    <au><snm>Group</snm><fnm>OCDW</fnm></au>
    <au><snm>Plessis</snm><fnm>L</fnm></au>
    <au><snm>Faria</snm><fnm>NR</fnm></au>
    <au><snm>Li</snm><fnm>R</fnm></au>
    <au><snm>Hanage</snm><fnm>WP</fnm></au>
    <au><snm>Brownstein</snm><fnm>JS</fnm></au>
    <au><snm>Layan</snm><fnm>M</fnm></au>
    <au><snm>Vespignani</snm><fnm>A</fnm></au>
    <au><snm>Tian</snm><fnm>H</fnm></au>
    <au><snm>Dye</snm><fnm>C</fnm></au>
    <au><snm>Pybus</snm><fnm>OG</fnm></au>
    <au><snm>Scarpino</snm><fnm>SV</fnm></au>
  </aug>
  <source>Science</source>
  <pubdate>2020</pubdate>
  <volume>497</volume>
  <fpage>493</fpage>
  <lpage>-497</lpage>
</bibl>

<bibl id="B11">
  <title><p>Sexuality, migration and AIDS in Ghana: a socio-behavioural
  study</p></title>
  <aug>
    <au><snm>Anarfi</snm><fnm>J. K.</fnm></au>
  </aug>
  <source>Health Transition Review</source>
  <pubdate>1993</pubdate>
  <volume>3</volume>
  <fpage>1</fpage>
  <lpage>-22</lpage>
</bibl>

<bibl id="B12">
  <title><p>Migration and AIDS</p></title>
  <aug>
    <au><snm>Decosas</snm><fnm>J.</fnm></au>
    <au><snm>Kane</snm><fnm>F.</fnm></au>
    <au><snm>Anarfi</snm><fnm>J.K.</fnm></au>
    <au><snm>Sodji</snm><fnm>K.D.</fnm></au>
    <au><snm>Wagner</snm><fnm>H.U.</fnm></au>
  </aug>
  <source>Lancet</source>
  <pubdate>1995</pubdate>
  <volume>346</volume>
  <issue>8978</issue>
  <fpage>826</fpage>
  <lpage>-828</lpage>
</bibl>

<bibl id="B13">
  <title><p>Population mobility and multi-partner sex in Botswana: implications
  for a spread of HIV/AIDS</p></title>
  <aug>
    <au><snm>Hope</snm><fnm>K.R.</fnm></au>
  </aug>
  <source>Journal of Reproductive Health</source>
  <pubdate>2001</pubdate>
  <volume>5</volume>
  <issue>3</issue>
  <fpage>73</fpage>
  <lpage>-93</lpage>
</bibl>

<bibl id="B14">
  <title><p>Factors in HIV/AIDS transmission in sub-Saharan Africa</p></title>
  <aug>
    <au><snm>Ateka</snm><fnm>G.K.</fnm></au>
  </aug>
  <source>Bulletin of the World Health Organization</source>
  <pubdate>2001</pubdate>
  <volume>79</volume>
  <issue>12</issue>
  <fpage>1168</fpage>
</bibl>

<bibl id="B15">
  <title><p>Labour Migration and HIV/AIDS in Southern Africa</p></title>
  <aug>
    <au><snm>Brummer</snm><fnm>D.</fnm></au>
  </aug>
  <source>IOM’s Regional Office for Southern Africa: Geneva</source>
  <pubdate>2002</pubdate>
</bibl>

<bibl id="B16">
  <title><p>International migration and the propagation of HIV in sub-Saharan
  Africa</p></title>
  <aug>
    <au><snm>Docquier</snm><fnm>F</fnm></au>
    <au><snm>Vasilakis</snm><fnm>C</fnm></au>
    <au><snm>Tamfutu Munsi</snm><fnm>D</fnm></au>
  </aug>
  <source>Journal of Health Economics</source>
  <pubdate>2014</pubdate>
  <volume>35</volume>
  <issue>8978</issue>
  <fpage>20</fpage>
  <lpage>-34</lpage>
</bibl>

<bibl id="B17">
  <title><p>Ranking the effectiveness of worldwide COVID-19 government
  interventions</p></title>
  <aug>
    <au><snm>Haug</snm><fnm>N</fnm></au>
    <au><snm>Geyrhofer</snm><fnm>L</fnm></au>
    <au><snm>Londei</snm><fnm>A</fnm></au>
    <au><snm>Dervic</snm><fnm>E</fnm></au>
    <au><snm>Desvars Larrive</snm><fnm>A</fnm></au>
    <au><snm>Loreto</snm><fnm>V</fnm></au>
    <au><snm>Beate Pinior</snm><fnm>B</fnm></au>
    <au><cnm>Stefan</cnm></au>
    <au><snm>Klimek</snm><fnm>P</fnm></au>
  </aug>
  <source>Nature Human Behaviour</source>
  <pubdate>2020</pubdate>
  <volume>4</volume>
  <fpage>1305</fpage>
  <lpage>-1312</lpage>
</bibl>

<bibl id="B18">
  <title><p>Cross-border mobility responses to COVID-19 in Europe: New evidence
  from Facebook data</p></title>
  <aug>
    <au><snm>Docquier</snm><fnm>F</fnm></au>
    <au><snm>Golenvaux</snm><fnm>N</fnm></au>
    <au><snm>Nijssen</snm><fnm>S</fnm></au>
    <au><snm>Schau</snm><fnm>P</fnm></au>
    <au><snm>Stips</snm><fnm>F</fnm></au>
  </aug>
  <source>Manuscript (Universit\'{e} catholique de Louvain)</source>
  <pubdate>2021</pubdate>
</bibl>

<bibl id="B19">
  <title><p>Bilateral international migration flow estimates for 200
  countries</p></title>
  <aug>
    <au><snm>Abel</snm><fnm>GJ</fnm></au>
    <au><snm>Cohen</snm><fnm>JE</fnm></au>
  </aug>
  <source>Scientific data</source>
  <publisher>Nature Publishing Group</publisher>
  <pubdate>2019</pubdate>
  <volume>6</volume>
  <issue>1</issue>
  <fpage>1</fpage>
  <lpage>-13</lpage>
</bibl>

<bibl id="B20">
  <title><p>Extensions of the {SEIR} model for the analysis of tailored social
  distancing and tracing approaches to cope with {COVID}-19</p></title>
  <aug>
    <au><snm>Grimm</snm><fnm>V</fnm></au>
    <au><snm>Mengel</snm><fnm>F</fnm></au>
    <au><snm>Schmidt</snm><fnm>M</fnm></au>
  </aug>
  <source>Scientific Reports</source>
  <pubdate>2021</pubdate>
  <volume>11</volume>
  <issue>1</issue>
  <fpage>4214</fpage>
  <url>https://www.nature.com/articles/s41598-021-83540-2</url>
  <note>Bandiera\_abtest: a Cc\_license\_type: cc\_by Cg\_type: Nature Research
  Journals Number: 1 Primary\_atype: Research Publisher: Nature Publishing
  Group Subject\_term: Epidemiology;Experimental evolution Subject\_term\_id:
  epidemiology;experimental-evolution</note>
</bibl>

<bibl id="B21">
  <title><p>Modeling and Analysis of Epidemic Diffusion with Population
  Migration</p></title>
  <aug>
    <au><snm>Liu</snm><fnm>M</fnm></au>
    <au><snm>Xiao</snm><fnm>Y</fnm></au>
  </aug>
  <source>Journal of Applied Mathematics</source>
  <pubdate>2013</pubdate>
  <volume>583648</volume>
  <fpage>8</fpage>
</bibl>

<bibl id="B22">
  <title><p>The introduction of population migration to SEIAR for COVID-19
  epidemic modeling with an efficient intervention strategy</p></title>
  <aug>
    <au><snm>Chen</snm><fnm>M</fnm></au>
    <au><snm>Miao</snm><fnm>L</fnm></au>
    <au><snm>Hao</snm><fnm>Y</fnm></au>
    <au><snm>Liu</snm><fnm>Z</fnm></au>
    <au><snm>Hu</snm><fnm>L</fnm></au>
    <au><snm>Wang</snm><fnm>L</fnm></au>
  </aug>
  <source>Information Fusion</source>
  <pubdate>2020</pubdate>
  <volume>64</volume>
  <fpage>252</fpage>
  <lpage>-258</lpage>
</bibl>

<bibl id="B23">
  <title><p>COVID-19 Crisis Management in Luxembourg: Insights from an
  Epidemionomic Approach</p></title>
  <aug>
    <au><snm>Burzynski</snm><fnm>M.</fnm></au>
    <au><snm>Machado</snm><fnm>J.</fnm></au>
    <au><snm>Aalto</snm><fnm>A.</fnm></au>
    <au><snm>Beine</snm><fnm>M.</fnm></au>
    <au><snm>Haas</snm><fnm>T.</fnm></au>
    <au><snm>Kemp</snm><fnm>F.</fnm></au>
    <au><snm>Magni</snm><fnm>S.</fnm></au>
    <au><snm>Mombaerts</snm><fnm>L.</fnm></au>
    <au><snm>Picard</snm><fnm>P.</fnm></au>
    <au><snm>Proverbio</snm><fnm>D.</fnm></au>
    <au><snm>Skupin</snm><fnm>A.</fnm></au>
    <au><snm>Docquier</snm><fnm>F.</fnm></au>
  </aug>
  <source>Economics and Human Biology</source>
  <pubdate>2021</pubdate>
  <volume>43</volume>
  <fpage>101051</fpage>
</bibl>

</refgrp>
} 









\clearpage
\end{backmatter}
\end{document}